\input amstex.tex
\input amsppt.sty
\magnification1200
\documentstyle{amsppt}
\NoBlackBoxes
\hcorrection{20 true mm}
\vcorrection{30 true mm}
\pagewidth{30 true pc}
\pageheight{47 true pc}
\define\lsim{\lesssim}

\document
\loadeurm
\loadbold

\topmatter
\title
Subquantum models:\\
basic principles, effects and tests
\endtitle

\rightheadtext{Subquantum models}

\author
Ji\v{r}\'i Sou\v{c}ek\\
\endauthor

\affil
\it Charles University, Prague
\endaffil

\address
\newline
Faculty of Mathematics and Physics
\newline
Sokolovsk\'a 83
\newline
186 00  Praha 8
\newline
Czech Republic
\medskip
\endaddress

\email
soucekj\@karlin.mff.cuni.cz
\endemail

\thanks
Preparation of this paper was supported by the~Grant No. RN\,19982003014
of~the~Ministry of~Education.
\endthanks

\abstract
We present models in which the indeterministic feature of~Quantum
Mechanics is represented in~the~form of~definite physical mechan\-isms. Our
way is completely different from so-called hidden parameter models,
namely, we start from a~certain variant of~QM -- deterministic QM -- which
has most features similar to~QM, but the evolution in~this theory is
deterministic. Then we introduce the~subquantum medium composed
of~so-called space-like objects. The~interaction of~a~deterministic
QM-particle with this medium is re\-presented by~the~random force, but it
is the~random force governed by~the~probability {\it amplitude} distribution.
This is the~quantum random force and it is very different from classical
random force. This implies that in our models there are no Bell`s
inequalities and~that our models (depending on~a~certain parameter
$\tau$) can be arbitrarily close to~QM. The~parameter $\tau$ defines
a~relaxation time and~on~time intervals shorter than $\tau$, the~evolution
violates Heisenberg`s uncertainty principle and~it is almost
deterministic -- spreading of~the~wave packet is much slower than in~QM.
Such type of~short-time effects form the bases of~proposed tests, which can,
in~principle, define limits of~validity of~QM. The proposed experiments
are related to~the~behavior of~quantum objects on~short time intervals,
where we expect the~behavior different from~QM. The~main proposed
feature is violation of~uncertainty relations on~short time intervals.

\quad

\quad

\quad

\quad

\quad

\quad

\endabstract


\pagebreak
\toc
\specialhead{} Introduction \endspecialhead
\head 1. Space-like objects and SLO-vacuum \endhead
\head 2. Deteministic Quantum Mechanics \endhead
\head 3. The Subquantum models SubQM and SubQM$_{RF}$ \endhead
\head 4. The unitary evolution in the model SubQM$_{RF}$ \endhead
\head 5. The interpretation of SubQM-models \endhead
\head 6. The long-time approximation and the Quantum mechanical limit
of~SubQM$_{RF}$ \endhead
\head 7. The short-time approximation and the concentration effect \endhead
\head 8. The correlated random force model SubQM$_{CRF}$ and~the~correlation
effect \endhead
\head 9. Proposed experiments \endhead
\specialhead{} Conclusions \endspecialhead
\endtoc
\endtopmatter


\newpage
\head
Introduction
\endhead

The main goal of~this research is to study the~indeterministic feature
of~the~Quantum Mechanics (QM). We construct models describing possible
mechanisms of~this indeterminism. We call them {\it subquantum models},
since
\newline
(i) these models contain more details with respect to~QM, but QM is
the~limit of~these models,
\newline
(ii) these models are part of~the~general quantum theory (defined
by~Feynman`s rules on~the~probability amplitude); they do not contain
any classical concept like the~probability distribution.

Our subquantum models are strictly different from so-called hidden
parameter models. The main difference is the following. If some model
contains certain random element, say something like a~random force, then
we postulate the~probability amplitude distribution of~these random
forces (and not the~probability distribution postulated in~the~hidden
parameter models). The reason is that we consider these random forces
as a~quantum phenomenon and~not as a~classical phenomenon. Our
subquantum models are based on~the~general quantum principle --
the~probability amplitude.

The hidden parameter models are, from our point of~view, inconsistent
mixtures of~quantum and classical concepts. They attempt to~explain
the~quantum phenomenon of~the~indeterminism by~using the~classical
concept of~the~probability distribution. This is impossible and Bell`s
inequalities prove that this mixture of~classical and~quantum concepts
is inconsistent. Our subquantum models contain no analogs of~Bell`s
inequalities.

A~rational model of~any indeterminism phenomena must contain two
elements:
\newline
(i) the underlying "deterministic" model,
\newline
(ii) the "random" elements which model the~indeterminism.

These two parts together must give a~model close to~QM. Both elements
must belong to~the general quantum theory.

The general quantum theory contains the following Feynman`s principles:
\roster
\item"(a)" to each possible trajectory of the system, there is associated
a~probability amplitude of~the~form $\Cal A~= \exp (iS / \hbar)$, where
$S$ is an~action associated to~this trajectory,
\item"(b)" for principially indistinguishable alternatives the~probability
amplitudes are summed,
\item"(c)" for principially distinguishable alternatives the~probabilities
$P = |\Cal A|^2$ are summed. The probability $P$ of~an~elementary
(principially distinguishable) event is given by $P = |\Cal A|^2$.
\endroster
The last Feynman`s postulate:
\roster
\item"(d)" all possible trajectories contribute with the same weight,
does not make part of~the~general quantum theory. This postulate is
a~basis of~the~indeterminism of~QM. We study mainly the~general quantum
models without this postulate.
\endroster

The "deterministic" QM model satisfies all postulates (a)--(c), but it
does not satisfy (d). In~this model the~system moves along (contrarily
to~(d)) classical trajectories but the~state is defined by~the~wave
function (more generally, by the~density matrix) and~standard quantum
rules (a)--(c) are applied. Only the~evolution operator is different from
QM, mainly the~wave packets do not disperse.

The random element is introduced into the~deterministic QM
\linebreak
by~postulating the~existence of~a~certain medium which we call {\it
SLO-vacuum}. This is a~medium composed from space-like objects.
Space-like objects are hypothetical new objects which are not directly
observable.

The main feature of~a~space-like object is its non-localizability.
The~trajectory of~a~freely moving space-like object is the~hyperplane in
$\Bbb R^4$ given by the~equation
$$
t = t_0 + \vec w . \vec x,
$$
where $\vec w = (w_1, w_2, w_3)$ (called the~{\it space-like velocity})
has the~physical dimension second/meter. We shall assume that
$$
|w| = \big( w_1^2 + w_2^2 + w_3^2 \big)^{1/2} < c^{-1},
$$
where $c$ is the velocity of~light.

The trajectory is the 3-dimensional hyperplane
$$
h_{\vec w, t_0} := \big\{ (t, \vec x) \in \Bbb R | \ t = t_0 + \vec w . \vec
x, \ \vec x \in \Bbb R^3 \big\}
$$
and the principal non-locality of~the~space-like object is clear. The
most typical space-like object is the~zero (space-like) velocity object
with the trajectory
$$
h_{0, t_0} := \big\{ (t_0, \vec x) | \ \vec x \in \Bbb R^3 \big\},
$$
i.e. the slice of~the space-time given by $t = t_0$, i.e. by fixing
time.

This feature makes space-like objects completely different from
standard (time-like or light-cone) objects and also from so-called
tachyons.

The main consequence of~non-localizability of~a~space-like object is
the~impossibility to~observe any particular space-like object.
Other\-wise, this does not imply that the system of ("infinitely") many
space-like objects cannot have observable consequences.

We suppose that there exists "vacuum" composed from space-like objects
-- the~SLO-vacuum. We assume that the DetQM-particles interact with this
SLO-vacuum. This is our basic general subquantum model. This model
describes the~physical bases of~our subquantum models.

To make this model mathematically more simple, we assume that
the~interaction of~DetQM-particles with space-like objects (from
SLO-vacuum) can be modelled by~the~concept of~the~random force. We
assume the~simplest possible probability amplitude distribution for
these random forces and this gives the basic SubQM$_{RF}$-model.

The basic subquantum model is parametrized by~a~certain constant which
defines how close the~given subquantum model is to~QM.
This basic constant of~our subquantum model can be interpreted as
a~relaxation constant $\beta = 1/\tau$, where $\tau$ is a~relaxation
time. The meaning of~$\tau$ is that on~time intervals $\Delta t \gg
\tau$ the~subquantum behavior approaches the~QM-behavior.

On the other hand, on short time intervals $\Delta t \ll \tau$,
the~subquantum effects can happen. In~particular, so-called concentrated
states can exist, which do not satisfy the~uncertainty relations. We can
prepare states for~which
$$
\Delta p \cdot \Delta x \ll \hbar.
$$
These concentration subquantum effects can exist and make
the~observational differences between subquantum models and~standard QM.

These subquantum effects can happen only under specific circumstances
described below.

We show that in the long-time ($\Delta t \gg \tau$) limit the~subquantum
models reduce to~QM, while in~the~short-time ($\Delta t \ll \tau$) limit
the~subquantum effects happen (the~concentration effect, the correlation
effect) under specific circumstances.

Our models represent the non-locality of~QM by the~concept
of~SLO-vacuum. This must be explained in~more details. The first thing
is to~note that concepts of~causality and~locality are completely
independent. It is possible to~reformulate the~electro-dynamics as
a~theory without electro-magnetic field but with a~force
acting-on-distance (this was done e.g. by~Feynman). Such a~theory is
non-local but causal. Non-locality is resolved by~introducing
an~electro-magnetic field.

Non-locality of~QM (consider EPR-pairs, the~teleportation etc.) is
clearly of~a~space-like character (and not of~light-cone character
typical for~electro-dynamics) and~it is completely natural that
the~resolution of~non-locality of~QM is given by~using space-like
objects.

For the exact prediction of~SubQM behavior it is necessary to~know
trajectories of~all space-like objects. But any particular space-like
object is not observable. Thus only the~(probability amplitude)
stochastic features of~space-like objects can be assumed.

A~certain fine point is that the~resulting effects of~SLO-vacuum are
well-observed. These are mainly the~indeterminism of~QM or,
equi\-valently, Feynman`s postulate (d), that any thajectory contributes
to~the~total amplitude. At this level, our subquantum models are
something like the~quantum mechanical Brownian motion. In fact, this
analogy is not correct: QM is the quantum analog of~the Brownian motion,
while subquantum models are analogs of~Ornstein-Uhlen\-beck stochastic
process (see [6]).

One can ask clearly: why any trajectory contributes to~the~resulting
probability amplitude? Our answer is clear: the~particle is subjected
to~the~random force (originated from the~interaction with space-like
objects) and this makes any trajectory possible.

Subquantum effects happen on~short time intervals where relaxation
phenomena are not yet realized. So the~proposed tests concern
the~behavior of~quantum particles during short time intervals.

A~typical effect is the concentration effect where the~uncertainty
relation is explicitly broken, or the~correlation effect which is
especially interesting, showing directly the~existence of~the~subquantum
medium.

In Section 1, we introduce the~concept of~a~space-like object. This
concept was originally introduced during the~study of~the~quaternionic
quantum theory of~tachyons ([1], [3], [5], [6]). One consequence of this
theory is the~classical approximation to~this quaternionic quantum
tachyon and~this is exactly our space-like object. Here we introduce
also the~simplest model for SLO-vacuum.

In Section 2, we introduce the concept of~the~deterministic QM. This means
to~change Feynman`s assumption (d) to~its opposite, that only classical
trajectories contribute to~the~total transition amplitude. A~rudimentary
form of~this idea was presented in~[4] and also in~[2] (and it is
implicitly assumed in~[6]). The~existence of~some type
of~the~deteministic QM is clearly necessary if we want to introduce
an~explicit random mechanism in~SubQM. The~random element (like
SLO-vacuum) must be working inside certain deterministic situation.
Of~course, other models are also possible, but we think, our DetQM has
a~certain mathematically appealing form.

In Section 3, the proper subquantum models are introduced. The general
model
$$
\text{SubQM := DetQM + SLO-vacuum}
$$
forms the physical basis. The~basic random model SubQM$_{RF}$ is
the~model where the~interaction of~the DetQM-particle with SLO-vacuum is
represented by~quantum random force. We call this random force as
quantum force since it is governed by~the~probability amplitude
distribution. This is the~main model studied in~this paper and it was
introduced in~[6].

In Section 4, we study the unitarity of~the~evolution of~the~state
in~SubQM$_{RF}$, especially the~simplest form of~the~relaxation
phenomena. We follow ideas from [6].

In Section 5, we study a~possible interpretation of~SubQM models from
the~point of~view of~QM-approximation. We observe that the~simplest
probability interpretation is inconsistent with the~right QM-limit.
The~second proposed interpretation seems to~be quite reason\-able. It
means that states with different (particle) velocities are principially
indistinguishable alternatives in~the~sense of~Feynman`s approach to~QM.
The~content of~this section is new, the~interpretation approved here was
implicitly assumed in~[6].

In Section 6, we study the long-time (with respect to the relaxation time
$\tau$) approximation to~SubQM$_{RF}$ and we show that, in~a~certain
sense, the~standard QM is obtained. The~passage
$$
\text{SubQM$_{RF}$ } \rightarrow \text{ QM}
$$
corresponds to the passage
$$
\text{Ornstein-Uhlenbeck } \rightarrow \text{ Brownian motion};
$$
this was especially made clear in~[6]. Here we do not repeat this
argument and~an~interested reader can find details in~[6]. The~ideas
of~this long-time approximation were already presented in~[6].

In Section 7, we study the~short-time approximation. We show that
on~short time intervals (relatively to~the~relaxation time $\tau$),
the~SubQM-behavior is completely different from QM-behavior. We study
quantitative details also by calculating spreading of~Gaussian wave
packets after passing through repeated slits.

We calculate quantitatively the breaking out of~uncertainty relations
and we define and study the~corresponding effect called {\it
the~concentration effect}. This concentration effect was implicitly (but not
quantitatively) mentioned in~[6] before studying the~more involved
"coherence effect".

Here we introduce the~important concepts: the particle`s momentum $\vec p$
and~the~QM-momentum $p^{(QM)}$. In~the~relaxed state the~mean value
of~$\vec p$ is infinite (like the~mean actual velocity of~a~Brownian
particle) and~only the~QM-momentum $p^{(QM)}$, defined by~the~Fourier
transform, can be used.

On the other hand, in~the~non-relaxed state (typically
in~the~concentrated state) the~mean value of~$p$ is finite (like
the~mean actual velocity of~an~Ornstein-Uhlenbeck particle can be finite
before approaching the~thermal equilibrium). We have only the~particle`s
momentum $p$, while the~QM-momentum is defined only for~the~relaxed
states. The~difference between $p$ and~$p^{(QM)}$ lies in~the~heart
of~the~subquantum models.

In Section 8, we introduce the~correlated random force model
\linebreak
SubQM$_{CRF}$, where the~random forces acting on~different particles are
inter-correlated. This is a~refined form of~SubQM$_{RF}$ and it was
introduced in~[6]. The~correlation effect was proposed in~[6] under
the~name "subquantum coherence effect". The~ideas can be found in~[6]
but without any calculations.

In Section 9, we propose possible experiments based
on~the~concentration effect or on~the~correlation effect. All proposed
experiments need to~study the~system during short time intervals.
Proposed physical values characterizing the~experiments are, in~certain
cases, calculated. The~main correlation experiment was already proposed
in~[6], but without any explicit calculation.

In Conclusions we discuss some general consequences of~subquantum models
and we also present our point of~view on~the~interpretation questions
of~QM (which were already briefly presented in~[6], where an~interested
reader can find some details not repeated here).

In preparing this paper I have taken advantage and~support from
discussion with many friends and colleagues. Among them, it is
a~pleasure for~me to~thank V.~Sou\v{c}ek, J.~Hrub\'y, M.~Giaquinta
and G.~Mo\-dica.

I also thank Eva Murtinov\'a for her invaluable help in typing of~this
manuscript.
            

\newpage
\head
1. Space-like objects and SLO-vacuum
\endhead

Here we show that assuming the Einsteinian locality but without
so-called "causality" assumption leads to new phenomena.

Dynamical state of an elementary object can be characterized \linebreak
by~the~energy-momentum vector $$(E, \vec p \,).$$

This vector lies in the Minkowski space and it can belong to one of
three possible types (assuming $c=1$):
\roster
\item"(i)" $(E, \vec p \,)$ is a time-like vector, $E^2 - {\vec p}^{\ \! 2} > 0$
    -- the object is a~time-like object, i.e. a~standard massive
    particle;
\item"(ii)" $(E, \vec p \,)$ lies on the light-cone, $E^2 - {\vec p}^{\ \! 2} = 0$
     -- the object is a~light-cone object, i.e. a~standard
     mass-less particle;
\item"(iii)" $(E, \vec p \,)$ is a space-like vector, $E^2 - {\vec p}^{\ \!2} < 0$
      -- the object is a~space-like object (SLO) and this is a new type of
      objects.
\endroster
\bigskip

Space-like objects are completely different from so-called
tachyons, as will be clearly seen below. Space-like object is
a~new concept of~an~"object" and it was
proposed in [1], [2].

The trajectory of the freely moving time-like object is given by
$$
\vec x = \vec x(t),\  t \in \Bbb R,
$$
with the constraint $|\dot{\vec x} (t)|^2 < c^2$, $\forall t \in \Bbb R$.

This constraint is a consequence of the tacit assumption of the
conservation of the type -- time-like, light-cone, space-like.
The velocity of~a~time-like object is given by
$$
\vec v(t) = \dot{\vec x} (t)
$$
and the constraint is $|\vec v|^2 < c^2$.

The acceleration is given by
$$\vec a (t) = \dot{\vec v} (t) = \ddot{\vec x} (t),\  t \in \Bbb R,$$
and Newton's law says
$$m.\vec a (t) = \vec F (\vec x (t),t).$$

Now for the space-like object we shall assume that its trajectory is
a~hypersurface
$$
t = f(\vec x), \  \vec x \in \Bbb R^3.
$$

This form is natural for the space-like object since it is lying locally
at each point outside the light-cone. We shall define the (space-like) velocity
of~a~space-like object as the local approximation by the gradient
$$\vec w~(\vec x) = \delta_{\vec x} f (\vec x) = \nabla_{\vec x} f (\vec
x), \  \vec x \in \Bbb R^3.$$

Clearly, the linear approximation has the form
$$t = \vec w \cdot \vec x + t_0.$$
This is the trajectory of a freely moving space-like object.

The condition to be outside the light-cone requires that
$$|\vec w~(\vec x)| < 1/c,\  \vec x \in \Bbb R^3$$
(remember that this $\vec w$ is the space-like velocity with the
physical dimension $(second)(meter)^{-1}$).

Then the acceleration matrix is given by
$$
\hat b(\vec x) = \delta_{\vec x} \vec w~(\vec x) =
\delta_{\vec x} \delta_{\vec x} f(\vec x) =
\big ( f_{x_\alpha x_\beta}\big )_{\alpha, \beta = 1}^3 \in M_{sym}^{3 \times 3}.
$$

It is a symmetric $3 \times 3$ matrix. The analogy of the Newton's law
for a~space-like object is the following
$$
m \hat b(\vec x) = m \delta_{\vec x} \delta_{\vec x} f(\vec x) =
\hat F(t(\vec x), \vec x),\ \vec x \in \Bbb R^3.
$$
Here, clearly, $\vec x$ is a "time variable" and it is a 3-dimensional
quantity.

This type of a theory with the multi-dimensional time was con\-sidered by
Carath\'eodory [7]. It follows that the force $\hat F(t(\vec x), \vec x)
\in M_{sym}^{3 \times 3}$ must satisfy the integrability condition
presented in~the~book of~Carath\'eodory.

The main consequence of these assumptions is non-locality
of \linebreak
a~space-like object. Typically, the trajectory of~a~freely moving
space-like object is
$$
t = \vec w \cdot \vec x + t^0
$$
and for a fixed $t = t_0$ we obtain the 2-dimensional plane (assuming
$\vec w~\not= 0$)
$$
\{ \vec x \in \Bbb R^3 | \ \vec w. \vec x = t_0 - t^0\} = \vec w \cdot
\frac{t_0 - t^0}{|\vec w|^2} + \vec w^{\bot},
$$
where $\vec w^{\bot} = \{\vec u| \ \vec u~. \vec w = 0\}$.

Such an object cannot be localized in a given laboratory. In fact, if
$L_{lab}$ is the linear estimation of the laboratory and $L_{cosm}$ is the
linear estimation of the cosmos, we obtain that only
$$(L_{lab}/L_{cosm})^2 \text{ - part of the space-like object}$$
is inside the laboratory.

From the fact that $L_{lab}/L_{cosm} \rightarrow 0$ we obtain that every
particular space-like object is non-localizable and then non-observable.

Such a conclusion allows us to make the following hypotheses:
\roster
\item"(i)" there exists large number of space-like objects,
\item"(ii)" the collective effect of all these space-like objects is observable.
\endroster

Of course, this means that a particular space-like object cannot be observed, but the
collection of a very large number (of order of $L_{cosm} / L_{lab}$)
of~them may be observable.

This allows us to assume the existence of many space-like objects. In other words:
non-observability of a single space-like object does not imply that the~system
of~many space-like objects cannot cause certain observable effects.

In fact, if there exist $L_{cosm} / L_{lab}$ space-like objects for each
second on~the~time axis, then there may exist observable collective effects
of~these space-like objects.

Later we shall show that the possible main collective effect of
SLO-vacuum is the non-deterministic behavior of particles in Quantum
Mechanics (QM).

We shall assume the following hypotheses. There exists freely moving
space-like object $O_\alpha$ for each $\alpha \in \Bbb Z$ ($\Bbb Z$ =
integers) such that each object $O_\alpha$ has a trajectory
$$
t = t_\alpha^0 + \vec {w_\alpha} . \vec x = f_\alpha (\vec x), \
\alpha \in \Bbb Z.
$$
We assume that parameters $t_\alpha^0$ and $\vec{w_\alpha}$ are randomly
distributed \linebreak in~the~following way.

Let $\tau_1 > \tau_0 > 0$ be the two fixed times and let for each
$\alpha \in \Bbb Z$, $\xi_\alpha$ be a random number in the interval
$(0,1)$ and $\vec {\eta_\alpha}$ be a random vector in the unit ball in
$\Bbb R^3$, $|\vec {\eta_\alpha}| < 1$.
We shall assume that $\xi_\alpha$ and $\vec {\eta_\alpha}$, $\alpha \in
\Bbb Z$, are independent random variables with the uniform distribution
in~$(0,1)$ and $\Bbb B^3$ (the~open unit ball in~$\Bbb R^3$), respectively.

Then we shall assume that for $\alpha \in \Bbb Z$
$$
\align
t_\alpha^0 &= \alpha.\tau_0 + \xi_\alpha.\tau_1,\\
\vec {w_\alpha} &= \vec {\eta_\alpha}.c^{-1}
\endalign
$$
where $c$ is the velocity of light (numerically $c=1$).

As a consequence we obtain that for a fixed $\vec x^0 \in \Bbb R^3$,
$t^0 \in \Bbb R$ we have the density of space-like objects at a~given $\vec x^0$
equal to
$$\lim_{T \rightarrow \infty} \frac{1}{T} \# \{\alpha | \ f_\alpha (\vec x^0) \in
(t_0, t_0 +T)\} = \frac{1}{\tau_0}$$
and this density does not depend on $t_0$.

Note that we could assume the general form of $O_\alpha$:
$$ t = f_\alpha (\vec x),$$
but we can simply assume the freely moving space-like objects, because
the~collective effect depends only very weakly on the detailed form of the
trajectories of space-like objects.

In the most simple form we can suppose that $\vec {w_\alpha} = 0$,
$\forall \alpha \in \Bbb Z$. (The~precise distribution of velocities
should be invariant with respect to~the~Lorentz group.) Nevertheless, this
simplest form is often sufficient.

The Newton's equations for the system of $n$ space-like objects with
trajectories $f^i (\vec x)$, $i = 1, \dots, n$ are given by
$$ m_i \delta_{\vec x} \delta_{\vec x} f^i (\vec x) =
 \hat F^i (f^1(\vec x), \dots, f^n(\vec x), \vec x)$$
or in the expanded form
$$ m_i \delta_{x_\alpha} \delta_{x_\beta} f(\vec x) = F_{\alpha \beta}^i
(f^1(\vec x), \dots, f^n(\vec x), \vec x)$$
$$ \alpha, \beta = 1,2,3,\ i = 1, \dots, n$$
where
$$\hat F^i(t^1, \dots, t^n, \vec x) = \big ( F_{\alpha \beta}^i(t^1, \dots, t^n,
\vec x)\big )_{\alpha, \beta} \in M_{sym}^{3 \times 3}.$$

It is clear that this system of equations is over-determined \linebreak
and~that there should exist well-defined integrability conditions.
\linebreak
The~method to~obtain these integrability conditions on $\hat F^i$ can be
found in~Carath\'eodory's book [7].


\newpage
\head
2. Deterministic Quantum Mechanics
\endhead

We shall consider in details the two Feynman principles mentioned
already in the Introduction:

(PA) Probability Amplitude,

(ID) Indeterminism.

The Probability Amplitude Principle says that for each observable event
there exists a~complex number $\Cal A_{event}$ such that the probability
of this event (being observed) is given by
$$\Cal P_{event}= |\Cal A_{event}|^2.$$

Let $(q_1, \dots, q_n)$ be generalized Lagrangian coordinates on the
configuration manifold $\bold M$.

For $t^1 < t^2$, $t^1, t^2 \in \Bbb R$, and $q^1, q^2 \in \bold M$, we
shall suppose that we have given the transition amplitude
$$\Cal A_{t^1, t^2} (q^1; q^2) \in \Bbb C$$
for the transition of the system from the initial state $q^1$ at the
time $t^1$ to the state $q^2$ at the final time $t^2$. The probability
of this transition is given by 
$$\Cal P_{t^1, t^2} (q^1; q^2) = |\Cal A_{t^1, t^2} (q^1; q^2)|^2.$$

Then we shall use the principle of the additivity of probability
amplitudes of the transition along different possible paths
$$\Cal A_{t^1, t^2} (q^1; q^2) = \sum \Cal A_{t^1, t^2} [q]$$
$$q(t^1)=q^1,\ q(t^2)=q^2, \tag{BC}$$
where (BC) are boundary conditions for the allowed trajectories and
$$\Cal A_{t^1, t^2} [q]$$
is the probability amplitude for a given trajectory $q(t)$, $t \in [t^1,
t^2]$. This probability amplitude is given by the formula using the
classical action
$$
\align
\Cal A_{t^1, t^2} [q] &= \exp\left\{ \frac{i}{\hbar} S_{t^1, t^2} [q]
\right\},\\
S_{t^1, t^2} [q] &= \int_{t^1}^{t^2} L(q, \dot q) dt.
\endalign
$$
Here $L(q, \dot q)$ is the Lagrange function of the system.

We shall reformulate all this on the phase space $\bold P$ with the
cannonical coordinates
$$ (q_1, \dots, q_n, p_1, \dots, p_n).$$
Here
$$p_i = \partial_{\dot{q_i}} L(q, \dot q), \  i=1, \dots, n,$$
and we express (assuming that this can be done) $\dot q$ as a~function
$$\dot{q_i} = \dot{q_i} (q,p).$$
The Hamilton function is then given by
$$H(q,p) = \sum \dot{q_i} (q,p) p_i - L(q, \dot q (q,p)).$$

For the trajectory
$$[q,p] = [q_1(t), \dots, q_n(t), p_1(t), \dots, p_n(t)]_{t=t^1}^{t^2}$$
in the phase space, the action can be expressed as
$$\bar S_{t^1, t^2}[q,p] = \int_{t^1}^{t^2}(\dot q(q,p).p -
H(q,p))dt,$$
where we use the bar $\bar S$ for the action expressed on the phase
space.

Then we have
$$
\align
\bar{\Cal A}_{t^1, t^2} [q,p] &= \exp \left\{ \frac{i}{\hbar}
\bar{S}_{t^1, t^2} [q,p] \right\},\\
\bar{\Cal A}_{t^1, t^2} (q^1; q^2) &= \sum_{q(t^1)=q^1, q(t^2)=q^2}
\bar{\Cal A}_{t^1, t^2} [q, p]
\endalign
$$
and
$$\bar{\Cal P}_{t^1, t^2} (q^1; q^2) = |\bar{\Cal A}_{t^1, t^2}
(q^1; q^2)|^2.$$

It is a well-known fact that this phase space approach gives \linebreak
the~same resulting transition amplitude as for the configuration approach
mentioned before, i.e.
$$ \bar{\Cal A}_{t^1, t^2} (q^1; q^2) = \Cal A_{t^1, t^2} (q^1; q^2).$$

In fact, the standard way to obtain the Feynman formulation of~QM
starts from the Schroedinger equation, its propagator is then expressed
using the Trotter's formula as a phase space Feynman integral. It is
then transformed by integrating out all momentum variables to the
configuration space Feynman integral.

Our starting point will be the Feynman formulation on the phase space.

On the phase space we can set (or, at least, we can try to set) finer
boundary conditions fixing also the momentum variables, i.e.
$$ \bar{\Cal A}_{t^1, t^2} (q^1, p^1; q^2, p^2) = \sum_{(BC)}
\bar{\Cal A}_{t^1, t^2} [q,p] $$
where boundary conditions are
$$ q(t^1) = q^1,\  p(t^1) = p^1,\  q(t^2) = q^2,\  p(t^2) = p^2. \tag{BC} $$

This refinement is the main point in what follows. This is fundamental
in considering wave functions.

The standard wave function at the time $t = t^1$ is a~complex function
$\psi (q; t^1)$ on the configuration space $\bold M$ normalized by the
condition
$$ \int_{\bold M} |\psi (q; t^1)|^2 dq = 1.$$
We can think of $\psi$ as a probability amplitude distribution
on~the~configuration space $\bold M$.
Using the transition amplitude $\Cal A$
we obtain the~evolution of a~PA-distribution $\psi$:
$$ \psi (q^2; t^2) = \int_{\bold M} \Cal A_{t^1, t^2} (q^1; q^2) \psi (q^1; t^1)
dq^1.$$

On the phase space we can consider similarly the PA-distribution (or
phase space wave function)
$$ \psi (q,p; t)$$
with the evolution
$$
\psi (q^2, p^2; t^2) = \int \bar{\Cal A}_{t^1, t^2} (q^1, p^1; q^2,
p^2) \psi (q^1, p^1; t^1)\frac{dq^1 dp^1}{\hbar}.
$$

Of course, there are degenerate cases, where this type of the
PA-distribution (wave function) depending on the position and momentum
variables is meaningless. This will be considered in more details below.

In fact, QM is a degenerate case in this sense, while SubQM models
introduced here are not degenerate.

The second Feynman principle -- we call it the Indeterminism --
the~ID-principle -- specifies which trajectories contribute to the
trans\-ition amplitude and what is the weight with which a given trajectory
contributes.

In fact, this specification was left in a complete dark above expressing
the total amplitude as a sum over trajectories. But one must specify
which trajectories are allowed and what are their weights.

There are many possible specifications and the Feynman \linebreak
ID-principle defines the extreme case by setting:
$$ \text {each trajectory contributes with the equal weight.}$$

The exact meaning of the phrase "with equal weight" is the content
of~many studies, but here we do not need to consider these details.

The second Feynman principle is expressed as a Feynman integral
$$
\bar {\Cal A}_{t^1, t^2} (q^1; q^2) = \int_{q(t^i) = q^i, i=1,2}
\bar {\Cal A}_{t^1, t^2} [q,p] . \prod_{t=t^1, i=1}^{t^2, n}
\frac{dq_i(t) \, dp_i(t)}{\hbar}.$$

These two principles together give the same transition amplitudes as
in~the cannonical QM.

The uncountable product $\prod$ over $t \in [t^1, t^2]$ expresses the
so-called Feynman measure, which exists only formally. We shall use
Feynman integrals in a formal sense.

Of course, on the phase space we can also write
$$
\bar {\Cal A}_{t^1, t^2} (q^1, p^1; q^2, p^2) =
\int_{\Sb
q(t^i) = q^i\\
p(t^i) = p^i\\
i=1,2
\endSb }
\bar {\Cal A}_{t^1, t^2} [q,p] . \prod_{t, i} \frac{dq_i(t) \,
dp_i(t)}{\hbar},$$
where the short notation of $\prod$ over $t, i$ means, as above, that
\linebreak
$t \in [t^1, t^2]$, $i=1, \dots, n$.

If for the amplitude $\bar {\Cal A}_{t^1, t^2} [q,p]$ the standard QM
holds, then
\linebreak
the~resulting transition amplitude
$\bar {\Cal A}_{t^1, t^2} (q^1, p^1; q^2, p^2)$ does not effectively
depend on~$p^1$ and~$p^2$. We then obtain the independence
of~the~PA-distribution $\psi (q,p; t)$ on~$p$ and thus we have to~return
to~the~PA-distribution $\psi (q;t)$, which does not depend on~the~momentum
variable $p$.

As a consequence of the ID-principle we shall obtain the standard
Feynman integral for the transition amplitude
$$
\Cal A_{t^1, t^2} (q^1; q^2) = \int_{q(t^i) = q^i, i=1,2}
\bar {\Cal A}_{t^1, t^2} [q,p] . \prod_{t, i} \frac{dq_i(t) \,
dp_i(t)}{\hbar}.
$$

This ID-principle implies then the Heisenberg commutation
\linebreak
relations, the uncertainty principle and so on.

The indeterminism of QM follows from the fact that each possible
trajectory contributes (with the equal weight) to the resulting
transition amplitude. This implies, for example, spreading out
of~wave packets.

In fact, both Feynman principles are very mysterious and both are
confirmed by an infinite amount of observational data.

The first PA-principle says that instead of summing up probabilities,
one has to sum up probability amplitudes and it forms the basis of all
interference phenomena in the quantum physics. This principle cannot be
explained, it can be only considered as an axiom.

We shall assume this PA-principle as a fundamental law and we shall
not try to modify it in any way.

The second Feynman principle, the indeterminacy, is less fundamental and
we shall look for modifications of it. Subquantum models considered in
this paper start with modifying the second Feynman principle.

Our first subquantum model, the Deterministic Quantum \linebreak
mechanics, denoted as DetQM, starts with assuming the PA-principle unchanged,
but the~second ID-principle to the opposite extreme case. Instead of~assuming
that all trajectories contribute to the transition amplitude, we
postulate that only a~classical trajectory (maybe, a~finite number of
classical trajectories) contributes to the transition amplitude.

So, instead of ID-principle we shall postulate the Det-principle.
$$
\spreadlines{-3pt}
\gather
\text {For the transition on the phase space: }\\
(q^1, p^1) \text { at } t^1 \rightarrow (q^2, p^2) \text { at } t^2,
\tag {Det}\\
\text { only the classical trajectory contributes.}
\endgather
$$
I.e., only such trajectory that is a solution of the Hamilton equations.

Let
$$
\align
q_i(t) &= q_i(q^1, p^1, t^1; t), \\
p_i(t) &= p_i(q^1, p^1, t^1; t)
\endalign
$$
be the classical evolution of the system in the phase space starting
from the point $(q^1, p^1)$ at the time $t^1$.

On the phase space the phase volume is conserved
$$\prod_i dq_i(t) dp_i(t) = \prod_i dq_i^1 dp_i^1, \ \forall t \in [t^1,
t^2].$$
We have inverse maps for $t > t^1$
$$
\align
q_i^1 &= \hat q_i (q(t), p(t), t; t^1),\\
p_i^1 &= \hat p_i (q(t), p(t), t; t^1)
\endalign
$$
which give the inverse map to the evolution map
$$ (q^1, p^1) \mapsto (q(t), p(t)), \  t \geq t^1.$$

Following the PA-principle we obtain the amplitude for the classical
trajectory
$$
\spreadlines{-4pt}
\multline
\bar {\Cal A}_{t^1, t^2} (q^1, p^1; q^2, p^2) =
\exp \left\{ \frac{i}{\hbar} \bar {S}_{t^1, t^2} (q^1, p^1; q^2, p^2) \right\} .\\
.\hbar \prod_i \delta (q_i(q^1, p^1, t^1; t^2) - q_i^2)
\delta (p_i(q^1, p^1, t^1; t^2) - p_i^2).
\endmultline
$$
Here
$$\bar {S}_{t^1, t^2} (q^1, p^1; q^2, p^2) = \int_{t^1}^{t^2} \big(
\sum_{i=1}^n p_i H_{p_i} - H \big) (q, p)dt$$
is the action along the classical trajectory.

At this moment it is completely natural to consider the amplitude
distribution
$$ \psi (q, p; t)$$
which will be called the wave function.

Then the evolution of the wave function is given by
$$ \psi(q^2, p^2; t^2) =\int \bar {\Cal A}_{t^1, t^2} (q^1, p^1; q^2,
p^2) \psi(q^1, p^1; t^1) \frac{dq^1 dp^1}{\hbar}.$$
It is now very important that the phase volume $dq\, dp$ is conserved
during the evolution.

This gives
$$
\psi(q^2, p^2; t^2) = \exp \left\{ \frac{i}{\hbar} S_{t^1, t^2} (q^1, p^1; q^2, p^2)
\right\}
\psi(q^1, p^1; t^1)
$$
where $q^1$ and $p^1$ are defined by
$$ q^1 = \hat q (q^2, p^2, t^2; t^1), \
p^1 = \hat p (q^2, p^2, t^2; t^1).$$

Here $q_i(t)$, $p_i(t)$ satisfy the Hamilton equations
$$
\align
\dot q_i(t) - H_{p_i} (q(t), p(t); t) &= 0,\\
\dot p_i(t) + H_{q_i} (q(t), p(t); t) &= 0.
\endalign
$$

Expanding the evolution into the Feynman integral we obtain
$$
\spreadlines{-5pt}
\multline
\psi (q^2, p^2; t^2) = \int \bar{\Cal A}_{t^1, t^2} [q,p] .
\prod_{t,i} \delta (\dot q_i(t) - H_{p_i}(t)) \delta (\dot p_i(t) +
H_{q_i}(t)) . \\
. \hbar^2 \prod_i \delta (q_i(t^1) - q_i^1) \delta (p_i(t^1) - p_i^1)
\delta (q_i(t^2) - q_i^2) \delta (p_i(t^2) - p_i^2) . \\
. \psi (q^1, p^1; t^1) \prod_{t, i} \frac{dq_i(t) dp_i(t)}{\hbar} .
\prod_i \frac{dq_i^1 dp_i^1}{\hbar}.
\endmultline
$$

It is clear that the~wave packets in DetQM do not spread out. If the~wave
function is supported in~the~certain phase volume, then the~evolved wave
function is supported in~the~region of~the~same volume.

We shall find the evolution equation by assuming the infinitesimal step
$\varepsilon \rightarrow 0$ in the time variable.
Let $t^2 = t^1 + \varepsilon$. Then
$$
\alignat2
&q_i^2 &= q_i^1 + \varepsilon \dot q_i &= q_i^1 + \varepsilon H_{p_i},\\
&p_i^2 &= p_i^1 + \varepsilon \dot p_i &= p_i^1 - \varepsilon H_{q_i},
\endalignat
$$
where all terms $o(\varepsilon)$ were neglected.

We also have
$$
S_{t^1, t^2} = \int_{t^1}^{t^2} (pH_p - H) dt = \varepsilon (pH_p -
H).
$$

For the infinitesimal evolution of the wave function we have
$$
\psi(q^2, p^2; t^2) = \psi(q^1, p^1; t^1) \left( 1 + \frac{i
\varepsilon}{\hbar} (pH_p - H) \right).
$$

On the other hand we have
$$
\psi(q^2, p^2; t^2) = \psi(q^1, p^1; t^1) + \sum_i \psi_{q_i}
\varepsilon H_{p_i} - \psi_{p_i} \varepsilon H_{q_i} + \varepsilon \psi_t
\big|_{q^1, p^1, t^1}.
$$

Then we obtain
$$
\psi + \psi_{q_i} \varepsilon H_{p_i} - \psi_{p_i} \varepsilon H_{q_i}
+ \varepsilon \psi_t = \psi + \frac{i \varepsilon}{\hbar} (pH_p - H) \psi.
$$

The first order terms give
$$
i\psi_t = \left( - i\sum H_{p_i} \partial_{q_i} + i\sum H_{q_i} \partial_{p_i}
- \frac{1}{\hbar} \sum p_i H_{p_i} + \frac{1}{\hbar} H \right) \psi. \tag{2.1}
$$

In the case where
$$
H = \sum \frac{1}{2m_i} p_i^2 + V(x), \ (x_i = q_i)
$$
we have
$$
H_{p_i} = \frac{1}{m_i} p_i, \  H_{q_i} = V_{x_i}.
$$

Then we obtain
$$
i \psi_t = \left( - \frac{1}{2} \sum \frac{1}{\hbar} \frac{1}{m_i} p_i^2 -
i \sum \frac{1}{m_i} p_i \partial_{x_i} + i\sum V_{x_i}
\partial_{p_i} + \frac{1}{\hbar} V \right) \psi. \tag{2.2}
$$
This is the Schroedinger equation in DetQM. This is a~genuinely quantum
equation, because there are terms depending on~$\hbar$. There is
a~principial difference between our equation and the~'t Hooft's equation
of~his deterministic QM [10]. Our equation gives non-dissipative quantum
evolution.
          
The general Schroedinger equation (2.1) can be written on
\linebreak
a~general differential manifold. The~analogical equation in~QM on~the general
manifold cannot be written without an~exact prescription of~the~order
of~operators. On~the~other hand, having given the~Hamilton function
$H(q,p)$, we can directly write the "Schroedinger-like" equation (2.1).


\newpage
\head
3. The subquantum models SubQM and SubQM$_{RF}$
\endhead

Now we shall apply the deterministic QM in the situation \linebreak
of~the~SLO-vacuum, i.e. the~medium composed of~many space-like objects.
We shall assume that there is a "sea" of space-like objects
as it was already assumed above at the end of Section 1.

We shall suppose that for each particle (time-like object) and each space-like object
there is a certain interaction. If $\psi (\vec x; t)$ is the state
\linebreak
of a time-like object and $t = f_\alpha (\vec x)$ is the trajectory of a
space-like object,
then the condition of the non-zero interaction between these two is
the following
$$
Spt \psi \cap \{ (t, \vec x) | \ t= f_\alpha (\vec x) \} \not= \emptyset.
$$
This implies that a given particle (described possibly by DetQM) interacts
with each space-like object.

At this moment, we are not able to specify the concrete form
of~the~interaction term. In fact, it is not necessary in this paper.

We shall call the system of DetQM for particles + SLO-vacuum
+ interaction between them as SubQM -- the general subquantum mechanical model.

We shall represent the total effect of the interaction particle --
space-like object -- as a random force acting on a particle. We shall denote this model SubQM$_{RF}$.
The random force depends on the point $(\vec x, t)$ and at this moment
we suppose the random forces at two different space-time points are independent
(components of the random force are also independent).

More concretely, we shall assume that the random force $F_i (t)$
acting on the $i$-th  particle (in fact, on the $i$-th degree of freedom)
at~the~time $t$ can be expressed as
$$
F_i (t) = \phi (F(t,x_i (t)), i, t)
$$
and that forces $F_i(t)$ and $F_j(t)$, $i \not= j$, are statistically
independent, because mostly $x_i(t) \not= x_j(t)$.

So that the general (ideal) model is 
$$ \text{SubQM } \approx \text{ DetQM + SLO-vacuum} $$
and the model with the random force representation of SLO-vacuum is
$$ \text{SubQM}_{RF}\  \approx \text{ DetQM + (SLO-vacuum)}_{RF}. $$

The last model consists of the following hypotheses. \newline
\roster
\item"(i)" There exists a random force $F_i(t)$ acting on the $i$-th degree of~freedom,
$i = 1, \dots, n$;
\item"(ii)" forces $F_i(t)$ and $F_j(t)$, $i \not= j$, are statistically
independent;
\item"(iii)" there is an amplitude distribution of the random force $F_i$ given by
$$
\Cal A_{t^1, t^2} [F_i] = \exp \left\{ i/\hbar \ \frac{a}{2}
\int_{t^1}^{t^2} F_i^2 (t) dt \right\} . \prod_{t \geq t^1}^{t^2} F_i(t)
$$
for each degree of freedom $i = 1, \dots, n$;
\item"(iv)" the system with $n$ degrees of freedom is described by DetQM
-- deterministic QM --  with a given random force.
\endroster
  
Assumption (iii) -- the amplitude distribution of a random force -- is
choosen as a simplest possibility.

Note that we assume the probability amplitude distribution \linebreak
$ \Cal A_{t^1, t^2} [F_i]$ and not the classical probability distribution.
This is a~consequence of the first Feynman principle (PA), which we consider
as a~general basis of quantum (and~subquantum) models.

From the independence assumption (ii) we obtain the join amplitude distribution for forces $F_1, \dots, F_n$ 
$$
\Cal A_{t^1, t^2} [F] = \exp \left\{ i/\hbar \ \frac{a}{2}
\sum_{i=1}^n \int_{t^1}^{t^2} F_i^2 (t) dt \right\} \prod_{t, i} dF_i(t),
$$
where $\prod_{t, i} dF_i(t)$ is "Feynman measure"
$$\prod_{t, i} dF_i(t) = \prod_{i=1}^n \prod_{t \in [t^1, t^2]} dF_i(t).$$

The system with $n$ degrees of freedom described by DetQM has
the transition amplitude given by formulas of the preceeding section
but with a change corresponding to the existence of a random force.
A~half part of Hamilton equations is changed to

$$ \dot p_i (t) = - H_{q_i} (q(t), p(t)) + F_i(t).$$
Then the transition amplitude is given by
$$
\gather
K_{t^1, t^2} (q^1, p^1; q^2, p^2) = \int \bar{\Cal A}_{t^1, t^2} [q,p] .
\prod_{t,i} \delta (\dot q_i - H_{p_i}) \delta (\dot p_i + H_{q_i} - F_i) . \\
\vspace{-8pt}
\ \  . \prod_i \delta (q_i(t^1) - q_i^1) \delta (p_i(t^1) - p_i^1)
\delta (q_i(t^2) - q_i^2) \delta (p_i(t^2) - p_i^2) . \tag{3.1}\\
\vspace{-5pt}
\ \ . \exp \left\{ i \frac{a}{2} \int_{t^1}^{t^2} \sum F_i^2 (t)dt \right\}
\prod_{t, i} dq_i(t) \prod_{t, i} dp_i(t) \prod_{t, i} dF_i(t).
\endgather
$$
For the special case when 
$$ H(x, p) = \sum \frac{1}{2m_i} p_i^2 + V(x) $$
we obtain after having done the integration on $\prod dp_i(t)$
$$
\gather
K~= \int \Cal A_{t^1, t^2} [x, \dot x] .
\prod_{t,i} \delta \big( m_i \ddot x_i(t) + V_{x_i}(x(t)) - F_i (t) \big). \\
\vspace{-9pt}
\ \ \ \ . (\text {BC}\delta xp) . \exp \left\{ i \frac{a}{2} \int \sum F_i^2 (t)dt
\right\} \prod_{t, i} dx_i(t) \prod_{t, i} dF_i(t), \ \ \ \
\tag{3.2}
\endgather
$$
where (BC$\delta xp$) denotes the boundary conditions term
$$ (\text{BC}\delta xp)\! :=\! \prod_i \delta (x_i(t^1) - x_i^1)
\delta (m_i \dot x_i(t^1) - p_i^1) \delta (x_i(t^2) - x_i^2) \delta (m_i \dot x_i(t^2) - p_i^2).
$$
Similarly, we denote
$$
(\text{BC}\delta qp) := \prod_i \delta (q_i(t^1) - q_i^1)
\delta (p_i(t^1) - p_i^1) \delta (q_i(t^2) - q_i^2) \delta (p_i(t^2) - p_i^2)$$
and, as a definition of the domain
$$
\spreadlines{-3pt}
\multline
(\text{BC}xp) := \{x_i (t^1) = x_i^1, m_i \dot x_i (t^1) = p_i^1,
x_i (t^2) = x_i^2, m_i \dot x_i (t^2) = p_i^2, \\
i=1, \dots, n \},
\endmultline
$$
$$
\spreadlines{-3pt}
\multline
(\text{BC}qp) := \{q_i (t^1) = q_i^1, p_i (t^1) = p_i^1,
q_i (t^2) = q_i^2, p_i (t^2) = p_i^2,\\
i=1, \dots, n \}.
\endmultline
$$

Integrating (3.2) with respect to $\prod dF_i(t)$ we obtain
$$
\gather
K=\int \Cal A_{t^1,t^2} [x, \dot x] . (\text{BC} \delta xp) .\\
\vspace{-9pt}
\ \ \ \ . \exp \left\{i \frac{a}{2} \int_{t^1}^{t^2} \sum \big( m_i \ddot x_i (t) +
V_{x_i} (x(t)) \big)^2 dt \right\} . \prod_{t, i} dx_i(t).\ \ \ \
\tag{3.3}
\endgather
$$

Then we obtain for the free evolution with
$$ V~\equiv 0, $$
$$
\gather
K_{t^1, t^2} (x^1, p^1; x^2, p^2) = \\
\vspace{-9pt}
= \int_{(BCxp)} \exp \left\{ \frac{i}{\hbar} \int_{t^1}^{t^2} \sum_i
\left( \frac{m_i}{2} \dot x_i^2 + \frac{a m_i^2}{2} \ddot x_i^2 \right) dt
\right\} \prod_{t, i} dx_i(t).
\tag{3.4}
\endgather
$$

The evolution of the wave function is then given by
$$
\psi(x^2, p^2; t^2) = \int K_{t^1, t^2} (x^1, p^1; x^2, p^2)
\psi(x^1, p^1; t^1) \prod_i dx_i^1 dp_i^1.
$$

Feynman integral in (3.4) is a Gaussian integral and it is separable for
$i=1, \dots, n$. It is then sufficient to calculate it for each degree
of~freedom separately. Thus we can suppose $n = 1$ and we have
$$
\spreadlines{-5pt}
\multline 
K_{t^1, t^2} (x^1, p^1; x^2, p^2) = \\
= \int_{(BCxp)} \exp \left\{ i \int_{t^1}^{t^2}
\left( \frac{m}{2} \dot x^2 (t) + \frac{a m^2}{2} \ddot x^2 (t) \right) dt \right\}
\prod_t dx(t).
\endmultline
$$

The result of the Gaussian integration has always the Gaussian form
$$
K_{t^1, t^2} = N_T \exp \left\{ \frac{i}{\hbar} \bar S_{t^1, t^2}
(x^1, p^1; x^2, p^2) \right\},
$$
where the normalization factor $N_T$ depends only on
$$ T = t^2 - t^1 $$
and $\bar S$ denotes the action calculated along the classical
trajectory (so-called classical action), i.e.
$$
\bar S_{t^1, t^2} (x^1, p^1; x^2, p^2) =  \int_{t^1}^{t^2} \frac{1}{2}
(m \dot{\bar x}^2 + a m^2 \ddot{\bar x}^2 ) dt
$$
where $\bar x (t)$ denotes the classical trajectory.

The classical action must satisfy the corresponding Euler`s \linebreak
equation
$$
- m \partial_t^2 \bar x (t) + a m^2 \partial_t^4 \bar x (t) = 0
$$
together with the boundary conditions
$$
(\overline {\roman {BC}}xp)\! := \{ \bar x (t^1) = x^1, m \dot {\bar x} (t^1) = p^1,
\bar x (t^2) = x^2, m \dot {\bar x} (t^2) = p^2 \}. \tag{3.5}
$$
Such boundary conditions are in a complete correspondence with the~fact
that Euler`s equation is a fourth order ordinary equation.

Our random force subquantum model contains a new constant $a$ with the
dimension $(time)^2 (mass)^{-1}$. It is useful to introduce the~relaxation
time $\tau$, resp. relaxation times $\tau_i$, by
$$
\tau^2 = am, \text{ resp. } \tau_i^2 = am_i,\  i = 1, \dots, n.
$$
The corresponding relaxation constants are
$$
\beta = 1/\tau, \text{ resp. } \beta_i = 1/\tau_i,\  i = 1, \dots, n.
$$

The meaning of the relaxation time $\tau$ is clear. \newline
(i) If we consider the time interval
$$ T = t^2 - t^1 \gg \tau,$$
then almost each trajectory contributes to the
trans\-ition amplitude.
\newline
(ii) If the considered time interval is small with respect to $\tau$:
$$
T = t^2 - t^1 \ll \tau,
$$
then only perturbed classical trajectories  contribute significantly
\linebreak
to~the~transition amplitude.

As a consequence we can expect that
\newline
(i) for $T \gg \tau$ the transition amplitude $K_{t^1, t^2}$ is close
to the standard quantum mechanical transition amplitude;
\newline
(ii) for $T \ll \tau$ the transition amplitude $K_{t^1, t^2}$ is close
to the deterministic transition amplitude from DetQM.

The free behaviour of a "SubQM - particle" (more exactly a SubQM degree
of freedom) is such that at short time intervals it is close to the~DetQM,
while on large intervals it is close to QM. The random force is
the cause of the transition from the DetQM region to the~QM region. This
transition needs some time, which is of order of~the~relaxation time
$\tau$. We shall study these limits -- SubQM and QM limits: $T \ll \tau$
and $T \gg \tau$ -- below in Sections 6, 7.

To calculate explicitly the transition amplitude we have to calculate at
first the classical trajectory $\bar x$ and then the value of the action
along this classical trajectory (classical action). Doing this
calculation for one degree of freedom, $n = 1$, we set $t^1 = -T$, $t^2 = T$,
$x^1 = -X$, $x^2 = X$ for $T>0$, $x \in \Bbb R$. The classical
trajectory satisfying boundary condition (3.5) is
$$
 \dot{\bar x} = a_1 \sinh \beta t + a_2 \cosh \beta t + a_0,
$$
where
$$
\align
a_1 &= \frac{\Delta v}{\sinh \beta T}, \\
a_2 &= \frac{\bar v~- V}{\cosh \beta T (1 - \tanh \beta T / \beta T)}, \\
a_0 &= \frac{V - \bar v~\tanh \beta T / \beta T}{1 - (\beta T)^{-1}
\tanh \beta T} \\
\endalign
$$
and
$$
\align
\Delta v~&= \frac{1}{2m} (p^2 - p^1), \\
\bar v~&= \frac{1}{2m} (p^2 + p^1), \\
V~&= \frac{x^2 - x^1}{t^2 - t^1}. \\
\endalign
$$
Then the classical action is
$$
\bar S_{t^1, t^2} (x^1, p^1; x^2, p^2) = \int_{t^1}^{t^2} \frac{m}{2}
\left( \dot{\bar x}^2 + am \ddot{\bar x}^2 \right) dt
$$
and the propagator
$$
K_{t^1, t^2} (x^1, p^1; x^2, p^2) = N_T \exp \left\{ \frac{i}{\hbar}
\bar S_{t^1, t^2} (x^1, p^1; x^2, p^2) \right\}.
$$
By doing the integration we obtain
$$
 \bar S~= \frac{m}{\beta} \Big[ (a_1^2 + a_2^2) \sinh \beta T \cosh
 \beta T + 2 a_0 a_2 \sinh \beta T + a_0^2 \beta T \Big]
$$
and then
$$
 \spreadlines{-6pt}
 \multline
 \bar S~= \frac{m}{\beta} \Big[ a_0^2 \beta T (1 - (\beta T)^{-1} \tanh
 \beta T) +\\
 + \left( \Big( a_2 + \frac{a_0}{\cosh \beta T} \Big)^2 + a_1^2
 \right) \sinh \beta T \cosh \beta T \Big].
 \endmultline
$$
Using the formula
$$
 a_2 + \frac{a_0}{\cosh \beta T} = \frac{\bar v}{\cosh \beta T}
$$
we obtain
$$
 \bar S~= \frac{m}{\beta} \left[ a_0^2 \beta T (1 - (\beta T)^{-1} \tanh
 \beta T) + \frac{\Delta v^2}{\tanh \beta T} + \bar v^2 \tanh \beta T \right]
$$
and finally
$$
 \bar S~= \frac{m (X - {\frac{\bar v}{\beta}} \tanh
 \beta T)^2}{T (1 - (\beta T)^{-1} \tanh \beta T)} + \frac{(p^1)^2
 + (p^2)^2}{2m \beta \tanh 2 \beta T} - \frac{p^1 p^2}{m \beta \sinh 2 \beta T}.
$$
The first term in this formula can be written as
$$
 \frac{m \left( x^2 - x^1 - {\frac{p^1 + p^2}{m \beta}} \tanh \beta T
 \right)^2}{4T \Big( 1 - (\beta T)^{-1} \tanh \beta T \Big)}.
$$
In the general case with $n \geq 1$ we have
$$
\spreadlines{-2pt}
\multline
\bar S_{t^1, t^2} (x^1, p^1; x^2, p^2) = \sum_{i=1}^n
\frac{1}{2\beta_i m_i} \left( \frac{(p_i^1)^2 + (p_i^2)^2}{\tanh \beta_i T}
- \frac{2 p_i^1 p_i^2}{\sinh \beta_i T} \right) + \\
+ \sum_i m_i \frac {\big[ x_i^2 - x_i^1 - (\beta_i m_i)^{-1}
(p_i^1 + p_i^2) \tanh (\beta_i \ T/2) \big]^2 }
{2T (1 - (\beta_i \ T/2)^{-1} \tanh (\beta_i \ T/2))},
\endmultline
$$
where
$$
T = t^2 - t^1.
$$
The normalization factor is given by
$$
N_T = (2 \pi i \hbar)^{-n} . \prod_i (\sinh \beta_i T)^{-1/2}
(\beta_i \ T - 2 \tanh \beta_i \ T/2)^{-1/2}.
$$
The simplest way to calculate $N_T$ is using the~well-known formula
for~quadratic Lagrangian $\Cal L$ (where $\vec x = (x, p)$),
$$
\multline
\int_{\vec x (t_1) = \vec x_1}^{\vec x (t_2) = \vec x_2} \exp \left\{
\frac{1}{\hbar} \int_{t_1}^{t_2} \Cal L (\vec x, \dot{\vec x}) dt \right\}
\prod_t dx(t) dp(t) = \\
= \frac{1}{2 \pi i \hbar} \left[ \det \left( - \frac{\partial^2
S_{Cl} (\vec x_2, \vec x_1)}{\partial \vec x_2 \partial \vec x_1} \right)
\right]^{1/2} \exp \left\{ \frac{i}{\hbar} S_{Cl} (\vec x_2, \vec x_1)
\right\}.
\endmultline
$$
Here $S_{Cl} (\vec x_2, \vec x_1)$ is the value of~the~action along
the~classical trajectory going from $\vec x_1$ to~$\vec x_2$.
The~determinant inside is the~well-known van Vleck-Pauli-Morette
determinant (see [9], formula (55)).

Now we shall calculate the evolution equation. It is possible to do it
by calculating all derivatives and then generalizing this to $n > 1$.
A~better way to prove this equation is presented in the next section.
The resulting equation for the wave function $\psi (x_1, p_1, \dots,
x_n, p_n; t)$ is
$$
i \hbar \partial_t \psi = \left( - \frac{1}{2} \sum_{i=1}^n
\frac{p_i^2}{m_i} - \frac{\hbar^2}{2} \sum_i m_i \beta_i^2
\partial_{p_i}^2 - i\hbar \sum_i \frac{p_i}{m_i} \partial_{x_i} \right)
\psi.
\tag{3.6}
$$

Now we shall introduce the interaction term into this equation. We shall
first consider its deterministic limit $\tau_i \rightarrow \infty$, i.e.
$\beta_i \rightarrow 0$. We obtain (setting $\hbar = 1$)
$$
i \partial_t \psi = \left( - \frac{1}{2} \sum \frac{p_i^2}{m_i} - i \sum
\frac{p_i}{m_i} \partial_{x_i} \right) \psi
$$
as a short-time limit of (3.5).

Comparison with the DetQM equation
$$
i \partial_t \psi = \left( - \frac{1}{2} \sum \frac{p_i^2}{m_i} - i \sum
\frac{p_i}{m_i} \partial_{x_i} + \sum V_{x_i} \partial_{p_i} + V~\right) \psi
$$
suggests the following evolution equation for the SubQM$_{RF}$ model
$$
\gather
i \hbar \partial_t \psi = \hat H \psi,\\
\vspace{3pt}
\hat H \psi = \\
\vspace{-16pt}
=\! \left( \! - \frac{1}{2} \sum_i \! \frac{p_i^2}{m_i} -
\frac{\hbar^2}{2} \sum_i \! m_i \beta_i \partial_{p_i}^2 -
i \hbar \! \sum_i \! \frac{p_i}{m_i} \partial_{x_i}
+ i \hbar \! \sum \! V_{x_i} \partial_{p_i} + V~\! \right) \! \psi.
\tag{3.7}
\endgather
$$
This is the basic equation of SubQM$_{RF}$. This is a~"Schroedinger-like"
equation for the wave function
$$
\psi (x_1, p_1, \dots, x_n, p_n; t),
$$
considered for each $t$ as an element of the Hilbert space
$$
\bold H = L^2 (\Bbb R_{(x)}^n \times \Bbb R_{(p)}^n ) \cong L^2 (\Bbb
R^{2n}).
$$
Then $\hat H$ is an operator defined on (a dense part of) $\bold H$ and
we see that $\hat H$ is formally Hermitian.
Thus the equation (3.6) generates the~unitary evolution in SubQM$_{RF}$.
Unitarity will be examined in~more details in the next section.

Let us note that in the general case of a manifold $\bold M$ the
corresponding Hilbert space will be the $L^2$-space on the cotangent
bundle $T^\ast \bold M$ and the evolution equation (3.6) makes a good
sense in this setting.


\newpage
\head
4. The unitary evolution in the model SubQM$_{RF}$
\endhead

In this section we shall study the evolution equation
$$
i \hbar \partial_t \psi = \hat H \psi, \ \hat H = \hat H_0 + \hat H_{int}
$$
where
$$
\hat H_0 = -\frac{1}{2} \sum_{i=1}^n \frac{p_i^2}{m_i} -
\frac{1}{2} \hbar^2 \sum m_i \beta_i^2 \partial_{p_i}^2 -
i \hbar \sum \frac{p_i}{m_i} \partial_{x_i},
$$
$$
\hat H_{int} = i \hbar \sum V_{x_i} (x) \partial_{p_i} + V(x).
$$

The wave function
$$
\psi (x_1, p_1, \dots, x_n, p_n;t)
$$
is a function on $\Bbb R_{(x)}^n \times \Bbb R_{(p)}^n  \cong \Bbb R^{2n}$
for each $t \in \Bbb R$.

From this equation we shall obtain the original Feynman integral. Then
we shall show in more details unitary properties of this evolution.

By Trotter`s formula we have (for $\varepsilon = T/m$)
$$
\gather
\langle x_0, p_0 | e^{-iHt} | x_m, p_m \rangle = \tag{4.1}\\
 = \lim_{m \rightarrow \infty} \int
\langle x_0, p_0 | e^{-i \hat H_{int} \varepsilon}
e^{-i \hat H_0 \varepsilon} | \xi_1, \eta_1 \rangle
\langle \xi_1, \eta_1 | x_1, p_1 \rangle \dots \\
\dots \langle x_{m-1}, p_{m-1} | e^{-i \hat H_{int} \varepsilon}
e^{-i \hat H_0 \varepsilon} | \xi_m, \eta_m \rangle
\langle \xi_m, \eta_m | x_m, p_m \rangle
\endgather
$$
where $\{ | x_k, p_k \rangle \}$ is $\delta$-basis of states, $x_k$,
$p_k \in \Bbb R^n$,
$$
| x_k, p_k \rangle \sim \delta_{x_k} (x) \delta_{p_k} (p)
$$
and $\{ | \xi_k, \eta_k \rangle \}$ is the dual basis
$$
| \xi_k, \eta_k \rangle  = \int e^{i \xi_k . x_k} e^{i \eta_k . p_k}
| x_k, p_k \rangle  d^n x_k d^n p_k
\sim e^{i \xi_k . x} e^{i \eta_k . p}.
$$

Here it is assumed the integration over all variables from the right
hand side (RHS) of equation (4.1) which do not enter the LHS of~this
equation. We shall obtain the standard product where we have \linebreak
written explicitly only the first term
$$
\spreadlines{-3pt}
\multline
\int \exp \Big\{ -i \varepsilon \Big[ - \frac{1}{2}
\sum \frac{p_{0i}^2}{m_i} + \frac{1}{2} \sum \beta_i^2 m_i \eta_{1i}^2 + \\
+ \sum \frac{p_{0i}}{m_i} \xi_{1i} -
\sum V_{x_i} (x_0) \eta_{1i} + V(x_0) \Big] - \\
- i \sum \xi_{1i} x_{0i} - i \sum \eta_{1i} p_{0i}
+ i \sum \xi_{1i} x_{1i} + i \sum \eta_{1i} p_{1i} \Big\} \dots
\endmultline
$$
Then we arrive at the term
$$
\spreadlines{-4pt}
\multline
\int \exp i \varepsilon \Big\{ \sum \xi_{1i} \frac{x_{1i} -
x_{0i}}{\varepsilon} + \eta_{1i} \frac{p_{1i} - p_{0i}}{\varepsilon} +
\frac{1}{2} \frac{p_{0i}^2}{m_i} - \\
- \frac{1}{2} \beta_i^2 m_i \eta_{1i}^2
- \frac{p_{0i}}{m_i} \xi_{1i} + V_{x_0} (x_0) \eta_{1i} + V(x_0) 
\Big\} \dots
\endmultline
$$
\smallskip

In the continuum limit $m \rightarrow \infty$ we obtain
$$
\spreadlines{-2pt}
\multline
\int_{(BC)} \exp i \Big\{ \int_{t^1}^{t^2} \sum \xi_i (t) \dot x_i (t)
+ \eta_i (t) \dot p_i (t) + \frac{1}{2m_i} p_i^2 (t) - \frac{1}{2} \beta_i
m_i \eta_i^2 (t) - \\
- \frac{p_i(t)}{m_i} \xi_i (t) + V_{x_i} (x(t)) \eta_i (t) - V(x(t)) \Big\}
dt \prod_{t,i} d\xi_i (t) d\eta_i (t) dp_i (t) dx_i (t).
\endmultline
$$

By integration with respect to $\prod d\xi_i (t) d\eta_i (t)$ we obtain
$\delta$-functions
$$
\spreadlines{-4pt}
\multline
\int_{(BC)} \prod_{t,i} \delta \left( \dot x_i (t) - \frac{p_i(t)}{m_i}
\right)
\exp i \Big\{ \int \frac{1}{2\beta_i^2 m_i} \Big( \dot p_i (t) +
V_{x_i} (x(t)) \Big)^2 dt + \\
+ \int \Big( \frac{1}{2} \frac{p_i^2 (t)}{m_i} - V(x(t)) \Big) dt \Big\}
\prod_{i,t} dp_i(t) dx_i(t).
\endmultline
$$

Using the preceding section we obtain here
$$
\align
\int \left( \frac{1}{2} \sum \frac{p_i^2}{m_i} - V~\right) dt &= \bar S,\\
\frac{1}{2\beta_i^2 m_i} &= \frac{a}{2}.
\endalign
$$

So that we have arrived at the initial Feynman integral. Integrating by
$\prod dp_i (t)$ we obtain
$$
\spreadlines{-4pt}
\multline
\int_{(BC)} \exp i \Big\{ \int \frac{1}{2\beta_i^2 m_i}
\Big( m_i \ddot x_i (t) + V_{x_i} (x(t)) \Big)^2 dt + \\
+ \int \Big( \frac{1}{2} m_i \dot x_i^2 (t) - V(x(t)) \Big) dt \Big\}
\prod_{t,i} dx_i(t).
\endmultline
$$

Now we shall study unitarity of the evolution in SubQM$_{RF}$. We shall
study namely the evolution in the $p$-space. We shall consider the case
of one degree of freedom
$$
\varphi (p^2, t^2) = \int K_{t^1, t^2} (p^1; p^2) \varphi (p^1, t^1) dp^1.
$$
Here
$$
K_{t^1, t^2} (p^1; p^2) = N_T^{(p)} \exp \big\{ i \bar S_{t^1, t^2}
(p^1; p^2) \big\}, \ T = t^2 - t^1
$$
and
$$
\bar S_{t^1, t^2} (p^1; p^2) = \frac{1}{2 \beta m} \left[
\frac{(p^1)^2 + (p^2)^2}{\tanh \beta T} - \frac{2 p^1 p^2}{\sinh \beta T}
\right].
$$
The reduced equation for $p$ is an "imaginary harmonic oscilator"
$$
i \partial_t \varphi = \hat H_0^{(p)}, \
\hat H_0^{(p)} := - \frac{1}{2} \hbar^2 m \beta^2 \partial_p^2 -
\frac{1}{2} \frac{p^2}{m}.
$$

The operator $- \hat H_0^{(p)}$ nor $\hat H_0^{(p)}$ is positive
definite, so that the~spectrum of $\hat H_0^{(p)}$ is not limited to a
half-line. This is rather a non-standard case and we shall show
explicitly that this operator creates a~unitary group of operators.
We shall also show that, in a certain sense, this unitary evolution has
certain stability (relaxation) properties.

Let $\{ T_t \}$, $T \in \Bbb R$, be a group of unitary transformations
defined for~$\varphi \in L^2 (\Bbb R)$ by
$$
(T_t \varphi) (x) := e^{- \beta t/2} \varphi (e^{- \beta t} x), \
x \in \Bbb R.
$$

Let Fourier transforms be defined as
$$
\align
(\Cal F \varphi) (p)&:= \int e^{-i/ \hbar x p} \varphi (x) dx,\\
(\Cal F^{-1} \varphi) (x)&:= \int e^{i/ \hbar x p} \varphi (p)
\frac{dp}{2\pi \hbar}.
\endalign
$$
Let $c_0 > 0$ be a fixed constant and let $\Cal U_0$ and $\Cal U_1$ be
two unitary transformations defined by
$$
\align
(\Cal U_0 \varphi) (x) &:= \exp \left\{ \frac{i}{\hbar}
\frac{c_0 \beta m}{4} x^2 \right\} \varphi (x),\\
(\Cal U_1 \varphi) (p) &:= \exp \left\{ \frac{i}{\hbar}
\frac{1}{2 c_0 \beta m} p^2 \right\} \varphi (p).
\endalign
$$

Then the following theorem holds (its proof was suggested to the author
by Dr. M. \v{S}ilhav\'y [8]).

Let us denote
$$
H_0^{(x)} = \hbar \beta \Big( \frac{i}{2} + i x \partial_x \Big).
$$

\proclaim{Theorem} \newline
(i) $T_t = e^{i/\hbar H_0^{(x)} t}$, i.e. $T_t$ is generated by
$H_0^{(x)}$.
\newline
(ii) Let $G = \Cal U_1 \Cal F \Cal U_0$. $G$ is a unitary operator.
If we denote
$$
H_1 = - \frac{1}{2} c_0 \hbar^2 \beta^2 m \partial_p^2 -
\frac{1}{2 c_0 m} p^2
$$
then $H_1$ is a unitary transformation of $H_0^{(x)}$ and
$$
H_1 = G H_0^{(x)} G^{-1}.
$$
Thus $H_1$ is a generator of a unitary group
$$
e^{i/\hbar H_1 t} = G T_t G^{-1}, \ t \in \Bbb R.
$$
\newline
(iii) We have an explicit form
$$
\spreadlines{-3pt}
\multline
\big( e^{- i/\hbar H_1 t} \varphi \big) (q) = (4 \hbar c_0 \beta m)^{-1/2}
(\sinh \beta t )^{-1/2} . \\
. \int \exp \left\{ \frac{i}{\hbar}
\frac{i}{2 c_0\beta m} \left[ \frac{p^2 + q^2}{\tanh \beta t} -
\frac{2pq}{\sinh \beta t} \right] \right\} \varphi (p) dp.
\endmultline
$$
\endproclaim

The proof uses the following lemma (due to M. \v{S}ilhav\'y, [8]).

\proclaim{Lemma}
We have
\newline
(i) $\partial_t T_t \big|_{t = 0} = \frac{i}{\hbar} H_0^{(x)},$
\newline
$$\text{(ii) } \Cal U_1 H_0^{(p)} \Cal U_1^{-1} = H_0^{(p)} + \frac{1}{c_0 m} p^2,
\text { where } H_0^{(p)} := \hbar \beta \big( \frac{i}{2}
+ ip \partial_p \big).\ \ \ \ \ \ \ \ $$
\newline
(iii) $\Cal {FH}_0^{(x)} \Cal F^{-1} = - H_0^{(p)}$, \
$\Cal F (x^2) \Cal F^{-1} = -\hbar^2 \partial_p^2.$
\newline
(iv) $\Cal U_0 \Cal H_0^{(x)} \Cal U_0^{-1} =  H_0^{(x)} +
\frac{1}{2}c_0 \beta^2 m^2 x^2.$
\newline
(v) $\Cal {FU}_0 H_0^{(x)} \Cal U_0^{-1} \Cal F^{-1} = - H_0^{(p)} -
\frac{1}{2} c_0 \hbar^2 \beta^2 m \partial_p^2.$
\newline
$$\text{(vi) } \Cal U_1 \partial_p^2 \Cal U_1^{-1} = \partial_p^2 -
\frac{p^2}{c_0^2 \hbar^2 \beta^2 m^2} - \frac{2}{c_0 \hbar^2 \beta^2 m}
H_0^{(p)}.\ \ \ \ \ \ \ \ \ \ \ \ \ \ \ \
\ \ \ \ \ \ \ \ \ \ \ \ \ \ \ \ \ \ \ \ $$
\endproclaim

The proof of the lemma is done by an explicit calculation. Proof of the
theorem follows as: \newline
(i) and (ii) by calculation using lemma. \newline
(iii) by calculation using the~formula
$$
\int e^{i a/2 p^2} e^{ipz} dp = \left( \frac{2 \pi}{a} \right)^{1/2}
e^{- \frac{i}{2a} z^2}. \qed
$$

For $\exp (- i H_1 t)$ we have obtained a formula consistent with
the~pro\-pa\-gator calculated before.

Now we shall study the transformed function
$$
\varphi (p;t)
$$
for large times $t \gg 1/\beta$. In this case we have
$$
\tanh \beta t \sim 1, \ \sinh \beta t \sim \frac{1}{2} e^{\beta t}.
$$
Let
$$
\varphi_1 (p) = \exp \left\{ i \frac{1}{2 \hbar \beta m} p^2 \right\}
\varphi (p)
$$
and let $\psi_1(q)$ be Fourier transform
$$
\psi_1(q) = \int \exp \left\{ i \frac{1}{\hbar \beta m} p q \right\}
\varphi (p) dp.
$$

If $\varphi_1$ is a compactly supported function then $\psi_1$ will be a
$C^{\infty}$- function. By the Theorem (ii) we obtain that the evolution
of $\varphi$ can be represented by $\psi_3$ defined as
$$
\align
\psi_2 (q;t) &:= (\sinh \beta t)^{-1} \psi_1
\left( \frac{q}{\sinh \beta t} \right), \\
\psi_3 (q;t) &:= \exp \left\{ - \frac{i}{\hbar\beta m} q^2 \right\}
\psi_2 (q;t).
\endalign
$$

We shall show that $\psi_2 (\cdot; t)$ relaxes to the "constant"
function \linebreak
for~$t \gg 1/\beta$. For $t \gg 1/\beta$ we have
$$
\psi_2(q;t) \approx e^{- \beta t/2} \psi_1 (q . e^{- \beta t}).
$$
The transformation $\psi_1 \mapsto \psi_2$ is a unitary transformation.
$$
  \int | \psi_2 (q;t)|^2 dq = \int |\psi_1|^2 dq = const.
$$

But for derivatives we have
$$
\partial_q \psi_2 \approx e^{- \beta t/2} . \psi_{1q} (q e^{- \beta t}).
e^{- \beta t}.
$$
Thus the $L^2$-norm of $D\psi_2$ goes to $0$:
$$
\int |D \psi_2 (q;t)|^2 dq= e^{- 2\beta t} \int |D \psi_1|^2 dq
\approx const.e^{- 2\beta t} \rightarrow 0.
$$

So that we have
$$
\psi_2(\cdot; t)\  "\rightharpoonup "\  const.
$$
In the same way we have
$$
\psi_3(\cdot; t)\  "\rightharpoonup "\  const. \exp \left\{ -
\frac{i}{\hbar \beta m} q^2 \right\}.
$$


\newpage
\head
5. The interpretation of SubQM-models
\endhead

Up to now, we have defined the states and the evolution in subquantum
models. The state of the system is defined by the wave function
$$
\psi = \psi (x,p), \ x, p \in \Bbb R^n.
$$
The wave function is considered as an amplitude distribution
of~position and momentum.

The evolution of the amplitude distribution in the momentum was studied
in the preceding section. We have found that the evolution
of~the~distribution $\varphi (p)$ is close to the "constant" distribution for
times much greater than the relaxation time. We call this type
of~$\varphi (p)$ {\it the relaxed distribution of momentum} and shall use
the name {\it relaxed region} for the situation when dependence on the
momentum $p$ in the wave function $\psi (x,p)$ is the relaxed
distribution.

In the relaxed region, a typical trajectory which contributes
significantly to the transition amplitude is highly irregular and in the
mathematical sense is non-differentiable. This means that the exact
velocity does not exist and that the mean velocity approaches infinity
on small time intervals. Such a behavior is typical for a Brownian
particle. The trajectory of the particle in the mathematical model of
Brownian motion is non-differentiable and the mean velocity is infinite.

By the analogy with the Brownian motion, we can suppose that a~typical
trajectory of the sub-quantum particle is similar to the typi\-cal
Brownian trajectory, when we have the relaxed situation. On the other
hand, this relaxed situation is achieved only approximately at~finite
times, so that the typical momentum is large but not infinite at~finite
times.

The conclusion is that it is very improbable that the dependence
of~$\psi (x,p)$ on the momentum $p$ could be observed. Thus we can
assume the rather conservative point of view that only the dependence of
the wave function $\psi (x,p)$ on the position variable $x$ can be
observed.
This leads to the following interpretation postulate.

We assume that in the subquantum models SubQM, SubQM$_{RF}$ and similar
models, where the state is specified by the wave function $\psi (x,p)$
depending on the position and momentum variables, the only observables
are operators depending only on $x$
$$
A~(x, \partial_x).
$$

The interpretation problem consists of two parts.
\newline
(i) If there is a 0-1 measurement (the~corresponding observable is
a~projection) then one has
to~define what will be the state after the measurement,
\newline
(ii) one has to define what is a probability of a positive outcome
of~this 0-1 measurement.
\newline
The first part (i) creates no interpretation problem.

Let $x \in \Bbb R_{(x)}^n$, $p \in \Bbb R_{(p)}^n$ so that the wave
function $\psi (x, p)$ is defined on the space
$$
\Bbb R_{(x, p)}^{2n} = \Bbb R_{(x)}^n \times \Bbb R_{(p)}^n.
$$
Then the state space is
$$
L^2 (\Bbb R_{(x, p)}^{2n}) \cong L^2 (\Bbb R_{(x)}^n) \otimes
L^2 (\Bbb R_{(p)}^n).
$$
$P$ is a projection in $L^2 (\Bbb R_{(x)}^n)$, i.e. the operator
$$
P: L^2 (\Bbb R_{(x)}^n) \rightarrow L^2 (\Bbb R_{(x)}^n)
$$
satisfying
$$
P^+ = P, \ P^2 = P, \ P \geq 0.
$$
Let $P \otimes id_{(p)}$ be the projection defined on
$L^2 (\Bbb R_{(x)}^n) \otimes L^2 (\Bbb R_{(p)}^n)$ by
$$
\psi \otimes \varphi \mapsto P\psi \otimes \varphi
$$
for $\psi \in R_{(x)}^n$, $\varphi \in R_{(p)}^n$, i.e. the projection
$P$ operates only on the variable $x$.

The interpretation postulate says that possible outcomes of the measurement
$P$ are 1 (answer = yes) or 0 (answer = no).
After the measurement the system will be in the state
$$
(P \otimes id_{(p)}) \psi
$$
if the result is 1, and in the state
$$
((id_{(x)} - P) \otimes id_{(p)}) \psi =
(id_{(x, p)} - P \otimes id_{(p)}) \psi
$$
if the result is 0.

The most typical situation is when
$$
\align
P &\cong \chi_{a, b}, \\
\chi_{a, b} &= \cases
                 1 \text{ for } a<x<b, \\
                 0 \text{ otherwise,}
                 \endcases
\endalign
$$
i.e. it is a characteristic function of the interval $(a, b)$. This
\linebreak
experiment is interpreted as passing through the "slit" $(a, b)$.
Then the operation of $P \otimes id_{(p)}$ is given by
$$
\psi (x, p) \mapsto \chi_{a, b} (x) \psi (x, p).
$$

On the other hand, we stated above that operators as $\chi_{a, b} (p)$
\linebreak
operating on the momentum variable are not observable.

At this moment it is necessary to make a clear distinction between the
particle`s momentum $p$ and the quantum-mechanical momentum (called
shortly QM-momentum) defined as follows.
The quantum-mechanical momentum is a property of the wave function, i.e.
of the amplitude distribution $\psi(x, p)$, and it is not a property of
a particle. The QM-momentum is defined by the behavior of the wave
function with respect to the translations in the physical space. The
corresponding momentum operator $- i \hbar \partial_{\vec x}$ is the
standard one. This does not depend on the $p$-distribution, since the
particle`s momentum $p$ is conserved by translations. Thus the wave
function with the~QM-momentum $\vec \xi$ is
$$
\psi (x, p) = const. e^{i \vec \xi . \vec x} \varphi (\vec p),
$$
where the dependence of $\varphi (\vec p)$ on $\vec p$ is a degeneracy
factor.

Thus concepts of the particle`s momentum and the QM-momentum are
completely different. Particle`s momentum is defined (not observable)
at each point of the particle`s trajectory while QM-mo\-men\-tum is
a~property of the function $\psi(\cdot, p)$. Both quantities have their
amplitude distribution but the amplitude distribution of the QM-momentum
is defined as a Fourier transform of $\psi(\cdot, p)$, the amplitude
distribution of particle`s momentum $\psi (x, \cdot)$ is independent
on~the~$x$-variable of $\psi$. In fact, in the relaxed region the
distribution in~$p$ approaches a~distribution of the~type
$$
\exp \left\{ i (2 \hbar \beta m)^{-1} p^2 \right\}.
$$

The second part of the interpretation -- the formula for probabilities
-- is less clear. In general, one can construct the density operator
corresponding to the pure state $\psi (x, p)$ as
$$
\rho (x, p; x', p') := \psi (x, p) \psi^* (x', p').
$$
The general density operator then will be
$$
\rho (x, p; x', p') = \sum_{i=1}^\infty \alpha_i \psi_i (x, p)
\psi_i^* (x', p'),\  \alpha_i \geq 0,\  \sum \alpha_i = 1.
$$

In the usual QM, the mean value of~the~observable $A(x, x')$ in the
state $\rho(x, x')$ is given by the formula
$$
Tr (A~\rho) = \int A(x, x') \rho(x, x') dx dx'.
$$
Now we have to treat variables $p$ and $p'$ in~$\rho (x, p; x', p')$
in~a~certain way.

In general, let us consider the~positive kernel $P (p, p')$. Let us
define the relative mean value of~$A$ as
$$
Tr_P (A~\rho) := \int A(x, x') \rho(x, p ;x', p') P (p, p') \, dx \, dp \,
dx' dp'
$$
and then the absolute mean value is given by
$$
Tr_P (A~\rho) / Tr_P (\rho).
$$
For the choice of~$P$ we have two main possibilities.

(IP$_1$). We set
$$
P (p, p') = \delta (p - p').
$$
This is formally the most standard (but wrong) choice. This means that
the~value $p$ is principially observable; in Feynman`s language it says
that the~alternatives with $p \not= p'$ are principially
distinguishable.

For the projection
$$
P_{a, b}: \psi (x, p) \rightarrow \chi_{a, b} (x) \psi (x, p)
$$
we obtain
$$
P (P_{a, b} \psi) = \int_a^b dx \int dp |\psi (x, p)|^2.
$$

In fact, this interpretation brings serious problems. There are
concrete problems with the QM-limit, which we shall show later.
This contradicts our point of~view we explained above.

(IP$_2$). We set
$$
P (p, p') \equiv 1 \text{ for all } p \text{ and } p'.
$$
The physical meaning of~this assumption is that situations with
different $p \not= p'$ are not principially distinguishable.
The meaning is that for different $p`$s we have to sum up
amplitudes, not probabilities. This corresponds to the view that
different $p$`s mean something like different positions at the times
$t_2$ and $t_2 - \varepsilon / 2$. Here $\varepsilon > 0$ denotes
the~time step.

This means to sum up amplitudes for different $p$ at first and then
to~do the square modulus. We have then
$$
P (P_{a, b} \psi ) = \int_a^b dx \int dp \, dp' \psi(x, p) \psi^* (x, p').
$$
In general we have
$$
P (\text{pos. outcome}) = \int \rho (x, p; x, p') dx \, dp \, dp'.
$$
Of course, in this case it is necessary to make a renormalization
of~the~wave function. The way we prefer is to consider the~probability
as a~relative probability, which has
to~be renormalized for each time moment in~such a~way that the total
probability be equal to~1.



\newpage
\head
6. The long time approximation and~the~quantum mechanical limit
of~SubQM$_{RF}$
\endhead

The long time approximation is defined by time intervals satisfying
$$
\Delta t \gg \frac{1}{\beta_i}, \ i = 1, \dots, n.
$$
In this approximation all exponential functions are very close to their
\linebreak
limits, i.e.
$$
\tanh \beta T \sim 1, \ \sinh \beta T \sim \infty.
$$
This type of approximation can be called also {\it the exponential
\linebreak
approximation}. The meaning is to set $\exp (\beta T) = \infty$
and~$\exp (- \beta T) = 0$.

The long time (exponential) approximation to the propagator from
Section~3 is the following.
$$
\multline
\bar S_{t^1, t^2} (x^1, p^1; x^2, p^2) =\\
= \sum_{i = 1}^n \frac{1}{2
\beta_i m_i} \left( (p_i^1)^2 + (p_i^2)^2 \right)
+ \sum_{i = 1}^n m_i \frac{[x_i^2 - x_i^1 - (\beta_i m_i)^{-1}
(p_i^1 + p_i^2)]^2}{2 T (1 - 2/(\beta_i T))}.
\endmultline
\tag{6.1}
$$
The use of the iterated propagator needs another representation --
separation of quantities into the groups:
\roster
\item"(i)" input quadratic form with the $2 \times 2$ matrix $Q_{inT}$,
\item"(ii)" output quadratic form with the $2 \times 2$ matrix $Q_{outT}$,
\item"(iii)" the in-out bilinear form with the $2 \times 2$ matrix $Q_{trT}$.
\endroster

Here
$$
\gather
\bar S~= a_T \frac{1}{2} \big({p^1}^2 + {p^2}^2\big) + b_T p^1 p^2 + c_T \frac{1}{2}
(x^2 - x^1)^2 + d_T (x^2 - x^1) (p^1 + p^2),\\
Q_{inT} = \left(\matrix
                 a_T \!\! &-d_T \\
                -d_T \!\! &\ c_T
                \endmatrix \right),\
Q_{outT}= \left(\matrix
                a_T \!\! &d_T \\
                d_T \!\! &c_T
                \endmatrix \right),\
Q_{trT}= \left(\matrix
                \ b_T \!\! &\ d_T \\
                -d_T \!\! &-c_T
                \endmatrix \right)
\tag{6.2}
\endgather
$$
and
$$
\align
a_T &= \frac{1}{\beta m} \left[ \frac{1}{\tanh \beta T} + \frac{tanh^2
(\beta T/2)}{\beta T - 2 \tanh (\beta T/2)} \right],\\
b_T &= \frac{1}{\beta m} \left[ - \frac{1}{\sinh \beta T} + \frac{tanh^2
(\beta T/2)}{\beta T - 2 \tanh (\beta T/2)} \right],\\
\vspace{-7pt}
& \tag{6.3}\\
\vspace{-7pt}
c_T &= \frac{\beta m}{\beta T - 2 \tanh (\beta T/2)},\\
d_T &= - \frac{\tanh (\beta T/2)}{\beta T - 2 \tanh (\beta T/2)}.
\endalign
$$
The propagator is equal to
$$
K_T = N_T \exp \frac{i}{\hbar} \left\{ \frac{1}{2} {X^1}^\top Q_{inT} X^1
+ {X^1}^\top Q_{trT} X^2 + \frac{1}{2} {X^2}^\top Q_{outT}^\top X^2 \right\},
\tag{6.4}
$$
where
$$
X^1 = \left( \matrix
             p^1\\
             x^1
             \endmatrix \right),\
X^2 = \left( \matrix
             p^2\\
             x^2
             \endmatrix \right).
$$
In the long time approximation we have
$$
a_T \approx \frac{1 + \omega_T}{\beta m},\ b_T \approx \frac{\omega_T}{\beta m},\
c_T \approx \beta m \omega_T,\ d_T \approx - \omega_T
\tag{6.5}
$$
where
$$
\omega_T = \frac{1}{\beta T - 2}.
$$
This gives in the long time approximation
$$
\gather
Q_{inT} = \left(\matrix
                \frac{1 + \omega_T}{\beta m} &\omega_T\\
                \omega_T &\beta m \omega_T
                \endmatrix \right),\
Q_{outT}= \left(\matrix
                \frac{1 + \omega_T}{\beta m} &- \omega_T\\
                - \omega_T &\beta m \omega_T
                \endmatrix \right),\\
Q_{trT}= \omega_T \left(\matrix
                        \frac{1}{\beta m} &- 1\\
                        1 &- \beta m
                        \endmatrix \right).
\tag{6.6}
\endgather
$$

It is important to study the evolution of~Gaussian wave packets. Let us
consider the~wave function at~the~time $t = t^1$ in~the~form
$$
\psi_1 (X^1) = \exp \frac{1}{\hbar} \left\{ - \frac{1}{2} {X^1}^\top A_1
X^1 + i B_1^\top X^1 \right\}
\tag{6.7}
$$
where $A_1$ is a given $2 \times 2$ matrix and $B_1$ is a~given
2-dimensional vector.

The evolution is given by
$$
\psi_2(p^2, x^2; t_2) = \int K_{t_2 - t_1} (x^1, p^1; x^2, p^2)
\psi_1(p^1, x^1; t_1) dp^1 dx^1.
$$
As a result we obtain
$$
\multline
\psi_2 (X^2; t_2) = const (T) . \exp \frac{i}{\hbar} \left\{ \frac{1}{2}
{X^2}^\top Q_{outT} X^2 \right\} .\\
. \exp \frac{1}{\hbar} \left\{ - \frac{1}{2} (Q_{trT} X^2 + B_1)^\top
(A_1 - i Q_{inT})^{-1} (Q_{trT} X^2 + B_1) \right\} .
\endmultline
\tag{6.8}
$$
The inverse matrix of a~$2 \times 2$ matrix $A$ can be calculated by
$$
A^{-1} = \frac{A^{adj}}{\det A}, \
\left( \matrix
       a &b\\
       c &d
       \endmatrix \right)^{adj} =
\left( \matrix
       d &-b\\
       -c &a
       \endmatrix \right).
\tag{6.9}
$$

Let us start with the case where
$$
A_1 = \left( \matrix
             - \frac{i}{\beta m} &0\\
             0 &0
             \endmatrix \right),
$$
i.e. the wave function $\psi_1$ depends on $p^1$ as $\exp i (p^1)^2 /
2 \beta m$. For $R = i A_1 + Q_{inT}$ we obtain
$$
R^{-1} = \frac{1}{2 \beta m} \left( \matrix
                                    (\beta m)^2 &- \beta m \\
                                    - \beta m & \frac{2}{\omega T} + 1
                                    \endmatrix \right)
$$
with $det R = 2 \omega_T$.
Then we obtain
$$
\align
Q_{outT} - Q_{trT}^\top R^{-1} Q_{trT} &= {\frac{1}{\beta m} \left(
\matrix
1 & 0\\
0 & 0
\endmatrix \right)},\\
R^{-1} Q_{trT} &= {\frac{1}{\beta m} \left( \matrix
                                            0 & 0\\
                                            1 & - \beta m
                                            \endmatrix \right)}.
\endalign
$$
Using this we obtain by the Gaussian integration
$$
\spreadlines{-3pt}
\multline
\psi_2 (X^2) = N_T \exp \left\{ \frac{i}{\hbar} \frac{1}{2} {X^2}^\top
\left( Q_{outT} - Q_{trT}^\top R^{-1} Q_{trT} \right) X^2 \right\} . \\
. \exp \left\{ - \frac{i}{\hbar} B_1^\top R^{-1} Q_{trT} X^2 \right\} .
\exp \left\{ - \frac{i}{\hbar} \frac{1}{2} B_1^\top R^{-1} B_1 \right\}.
\endmultline
$$
In this way we obtain for $B_1 = \left( \matrix
                                        l^1\\
                                        k^1
                                        \endmatrix \right)$,
$$
\spreadlines{-2pt}
\multline
\psi_2 (p^2, x^2) = N_T \exp \left\{ \frac{i}{\hbar} \frac{1}{2 \beta m}
(p^2)^2 \right\} \exp \left\{ \frac{i}{\hbar} k^1 \left( x^2 -
\frac{p^2}{\beta m} \right) \right\} . \\
. \exp \left\{ - \frac{i}{4 \hbar \beta m} \left( \beta m l^1 - k^1
\right)^2 - \frac{i (k^1)^2}{2 \hbar \beta m \omega_T} \right\}.
\endmultline
$$
The absolute phase factor is equal to
$$
\exp \left\{ - \frac{i}{\hbar} \frac{(k^1)^2}{2m} T \right\} .
\exp \left\{ \frac{i}{4 \hbar \beta m} \left[ 4 (k^1)^2 - (\beta m l^1 -
k^1)^2 \right] \right\}.
$$
The first factor is exactly the QM-factor $\exp \{ - i E t \}$, while
the second factor is a~correction, which is a~constant independent
on~$T$.
In fact, in $\psi_1$ we have the~QM-momentum
$$
\psi_1 \approx \exp \left\{ \frac{i}{\hbar} \left( l^1 p^1 + k^1 x^1 \right)
\right\}.
$$
The term
$$
\psi_2 \approx \exp \left\{ \frac{i}{\hbar} \frac{1}{\beta m} \frac{1}{2}
(p^2)^2 \right\}
$$
is the stabilized particle momentum distribution which is already
in~the~propagator.

The term
$$
\psi_2 \approx \exp \left\{ \frac{i}{\hbar} k^1 x^2 \right\}
$$
is the standard conservation of~the~QM-momentum $k^1$, i.e. $k^2 = k^1$.

A~completely new feature is the last term
$$
\psi_2 \approx \exp \left\{ - \frac{i}{\hbar} \frac{1}{\beta m} k^1 p^1 \right\}.
$$
This means that in the long time approximation the dependence $\psi_1
\approx \exp \{ i l^1 p^1 \}$ on $p^1$ is forgotten, the value $l^1$ enters
only the~time-independent absolute phase factor mentioned above. On the other
hand, the dependence of $\psi_2$ on $p^2$ is governed by $k^1 / \beta
m$, i.e. it depends on the QM-momentum.

Let us now suppose that $\psi_1$ is the superposition of waves with $l^1
\equiv 0$ and different $k^1$:
$$
\psi_1 (p^1, x^1) \approx \int dk^1 \exp \left\{ \frac{i (p^1)^2}{2
\hbar \beta m} \right\} \exp \left\{ \frac{i}{\hbar} k^1 x^1 \right\}
a (k^1).
$$
Then by the superpositon principle we have
$$
\multline
\psi_2 \approx N_T \exp \left\{ \frac{i (p^2)^2}{2 \hbar \beta m}
\right\} .\\
. \int \exp \left\{- \frac{i (k^1)^2}{2 \hbar m} \left( T - \frac{3}{2 \beta}
\right) \right\} \exp \left\{\frac{i}{\hbar} k^1 \left( x^2 -
\frac{p^2}{\beta m} \right) \right\} a (k^1) dk^1.
\endmultline
$$
The difference between $t - 3 / (2 \beta)$ and $T$ is small for $\beta T
\gg 1$. But dependence on $x^2 - p^2 / (\beta m)$ is crucial.

Let us consider the simplest projector (the slit)
$$
\chi_{a,b} (x^2).
$$
Applying it we obtain $\varphi_2$:
$$
\varphi_2 (p^2, x^2) = \psi_2 (p^2, x^2) . \chi_{a,b} (x^2).
$$
Using the interpretation postulate (IP$_1$) we obtain the probability
$$
\spreadlines{-2pt}
\multline
Prob (a,b) = \int \psi_2 (p^2, x^2) \psi_2^* (p^2, x^2) \chi_{a,b} (x^2)
dp^2 dx^2 =\\
= \int a (k^1) a^* ({k^1}') \chi_{a,b} (x^2) \dots dk^1 d{k^1}'
dx^2 . \\
. \int \exp \left\{ - \frac{i}{\hbar \beta m} (k^1 - {k^1}') p^2
\right\} dp^2.
\endmultline
$$
The last integral gives $\delta (k^1 - {k^1}')$ and then
$$
Prob (a,b) = \int \left| a (k^1) \right|^2 \chi_{a,b} (x^2) dk^1 dx^2
$$
and this is clearly an incorrect result, in which the interference terms
are neglected. We think that this excludes the probability
interpretation (IP$_1$).

On the other hand, the probability formula in (IP$_2$)
$$
Prob (a,b) = \int \psi_2 (p^2, x^2) \psi_2^* ({p^2}', x^2) \chi_{a,b} (x^2)
dp^2 d{p^2}' dx^2
$$
is quite consistent.

Putting inside the formula for $\psi_2 (p^2, x^2)$ using the formula
$$
\int \exp \left\{ - \frac{i}{\hbar \beta m} \left( - \frac{(p^2)^2}{2} \right)
\right\} \exp \left\{ - i p^2 \frac{k^1}{\hbar \beta m} \right\} dp^2
= \exp \left\{ - \frac{i (k^1)^2}{2 \hbar \beta m} \right\}
$$
and calculating the integration with respect to $p^2$ and ${p^2}'$ we
obtain
$$
\spreadlines{-2pt}
\multline
Prob (a, b) = N_T^2 \int \exp \left\{ i \frac{(p^2)^2 - ({p^2}')^2}{2 \hbar \beta m}
\right\} . \\
. \exp \left\{ - i \frac{(k^1)^2 - ({k^1}')^2}{2 \hbar m}
\left( T - \frac{3}{2 \beta} \right) \right\}
\exp \left\{ i \frac{x^2 (k^1 - {k^1}')}{\hbar} \right\} . \\
. \exp \left\{ - i \frac{k^1 p^1 - {k^1}' {p^2}'}{\hbar \beta m} \right\}
\chi_{a, b} (x^2) a(k^1) a^* ({k^1}') dk^1 d{k^1}' dp^2 d{p^2}' dx^2 \approx \\
\approx
\int \exp \left\{ - i \frac{(k^1)^2 - ({k^1}')^2}{2 \hbar m}
\left( T - \frac{1}{2 \beta} \right) \right\} . \\
. \exp \left\{ i \frac{x^2 (k^1 - {k^1}')}{\hbar} \right\}
. \chi_{a, b} (x^2) a(k^1) a^* ({k^1}') dk^1 d{k^1}' dx^2.
\endmultline
$$
Now the formula is quite close to the QM formula. If we change
$T - (2 \beta)^{-1} \rightarrow T$ then we obtain exactly the~QM evolution.

In fact, in QM we have
$$
\psi_2 (x^2) = \int \exp\left\{ - \frac{i}{\hbar} \frac{(k^1)^2}{2m} T \right\}
\exp \left\{ i \frac{k^1 x^2}{\hbar} \right\} a (k^1) dk^1
$$
and this implies the QM probability formula
$$
\spreadlines{-3pt}
\multline
\int |\psi_2|^2 \chi_{a, b} dx^2 =
\int \exp \left\{ - i \frac{(k^1)^2 - ({k^1}')^2}{2 \hbar m} T \right\} . \\
. \exp \left\{ i \frac{(k^1 - {k^1}') x^2}{\hbar} \right\} a (k') a^* ({k^1}')
dk^1 d{k^1}'.
\endmultline
$$


\newpage
\head
7. The short-time approximation and the concentration effect
\endhead

In this section we shall consider in details what evolution can happen
during the~short-time intervals satisfying
$$
T \ll 1 / \beta.
$$
In these intervals the evolution is, in a~certain sense, close to
\linebreak
the~DetQM evolution.

We shall proceed in the following steps:
\roster
\item"(A)" the short-time approximation to the propagator,
\item"(B)" the concept of the concentrated state,
\item"(C)" the preparation of concentrated states,
\item"(D)" the short-time evolution of concentrated states,
\item"(E)" some exact calculations.
\endroster

In the preceding section we have found that after a~(free) long-time
evolution, the dependence of the wave function on the particle momentum
$p$ is of the type
$$
\psi (p, \cdot) \approx \exp \frac{i}{\hbar} \left\{ \frac{1}{2 \beta m}
p^2 \right\} \dots,
$$
i.e. all particle momenta $p$ contribute almost equally. We have called
these states relaxed states and their behavior is closed
to~the~QM-behavior.

The concentrated states are characterized as states where particle
momenta are localized, i.e. close to a~certain value $p_0$. These states
are very far from the~QM-states and at these states the main differences
between QM and subquantum models are presented -- and this is the main
theme of~the~rest of~the~paper.

A~typical concentrated state is of~the~form
$$
\psi (p, \cdot) \approx \exp \left\{ - \frac{1}{2} \frac{(p -
p_0)^2}{\Delta p^2} \right\} \dots
$$
The first question is if it is possible to prepare concentrated states.
It is possible by letting particles to pass through iterated slits.

The second question is to describe the non-QM behavior of~the~short-
time evolution of~concentrated states. The main feature is that the
dispersion of~the~wave packets is slower than the QM-dispersion and that
in~short-time evolution the original particle momentum $p_0$ is partially
remembered. After longer time, the memory of~$p_0$ is almost forgotten
and~the~relaxation happens.

(A) The short-time propagator is obtained by using the approximation for
$\beta T \ll 1$.
$$
\align
 \tanh \beta T &\approx \beta T - \frac{1}{3} (\beta T)^3 \approx \beta T \\
 \sinh \beta T &\approx \beta T + \frac{1}{6} (\beta T)^3 \approx \beta T \\
 \cosh \beta T &\approx 1 + \frac{1}{2} (\beta T)^2 \\
 1 - (\beta T / 2)^{-1} \tanh (\beta T / 2) &\approx \frac{1}{12} (\beta
T)^2 \\
 \tanh (\beta T / 2) &\approx \frac{1}{2} \beta T - \frac{1}{24} (\beta
T)^3 \approx \frac{1}{2} \beta T.
\endalign
$$

We obtain the short-time approximation
$$
\bar S~= \frac{1}{2 \beta m} \frac{(p^2 - p^1)^2}{\beta T} + \frac{\beta
m}{2} \frac{12}{(\beta T)^3} \left[ x^2 - x^1 - \frac{1}{m} \frac{p^1 +
p^2}{2} T \right]^2. \tag{7.1}
$$
Then $K = \exp \left\{ (i / \hbar ) \bar S~\right\}$.

We see that for short times, $\beta T \ll 1$, the dispersion of the wave
packets, determined mainly by the~second term, is similar to the~QM but
for $T (\beta T)^2$ instead of~$T$.

The first term is analogical to the~$x$-propagator in~QM, so that
the~short-time dispersion of the $p$-packet is similar to the~QM-dis\-persion
of~the~wave-packets.
In the initial period, where particle momenta $p^1$ and $p^2$ are
localized around $p_0$, the~$x$-dispersion is very slow because of~the~term
$(\beta T)^3$. Then the evolved distribution in $x_2$ is localized
around
$$
x_0^1 + \frac{1}{m} p_0^1 T,
$$
where $x_0^1$ is the center of the initial wave packet.
All these observations will be made more precise in what follows.

(B) The definition of the concentrated state is based on the idea
of~the~localization such that there is a~term
$$
\bar S~\approx - \frac{1}{2} \frac{(p - p_0)^2}{\Delta p^2} + \dots, \
\psi \approx \exp \left\{ \frac{i}{\hbar} \bar S~\right\}
$$
in the wave function.

Thus we said that the wave function $\psi (p, x)$ describes
the~concentrated state with the center $p_0$ and~the~dispersion $\Delta
p$ if there exists a~constant $c$ such that
$$
|\psi (p, x)| \leq c . \exp \left\{ - \frac{1}{2} \frac{(p -
p_0)^2}{\Delta p^2} \right\}. \tag{7.2}
$$
If the wave function can be written in~the~form
$$
\psi (p, x) = e^{i S_0 (p, x)} . e^{- S_1 (p, x)}, \ S_0 (p, x), \
S_1 (p, x) \in \Bbb R,
$$
then this condition means that
$$
c \cdot \frac{(p - p_0)^2}{2 \Delta p^2} \leq S_1 (p, x), \ \forall p, x \in \Bbb R.
\tag{7.3}
$$

Let us note that the concentrated state has nothing in common with
QM-states with the localized QM-momentum. The~QM-state with the
localized QM-momentum depends only on~$x$ and~its SubQM-approximation is
the state with the relaxed dependence
on~the~particle momentum $p$.

One can also consider the simultaneous localization in~$x$, so that the
completely concentrated state satisfies
$$
|\psi (p, x)| \leq c . \exp \left\{ - \frac{1}{2} \frac{(p -
p_0)^2}{\Delta p^2} \right\}  \exp \left\{ - \frac{1}{2} \frac{(x -
x_0)^2}{\Delta x^2} \right\}. \tag{7.4}
$$
We look for concentrated states with
$$
\Delta p \cdot \Delta x \ll \hbar / 2 \tag{7.5}
$$
which, in a sense, break the Heisenberg principle.

We shall characterize such states by~the~"degree" of~concentration
$\kappa$ defined as
$$
\kappa := \frac{2}{\hbar} \cdot \Delta x \cdot \Delta p.
$$
Concentration states are states satisfying
$$
\kappa < 1.
$$
 
In this case we have still the Heisenberg incertainty relation
$$
\Delta p^{(QM)} \cdot \Delta x \geq \hbar / 2,
$$
because this is a mathematical property of~the~Fourier transform. This
relation does not depend on any physics, only on~the~definition
\linebreak
of~the~QM-momentum. If we define the~QM-momentum through Fourier
transform, then the Heisenberg relation expresses the~mathematical
property of~this object. The physics lies in~defining the~QM-momentum
in~this way.

With respect to the particle momentum $p$, the situation with Heisenberg
relation is the following. In the relaxed state the particle momentum is
completely dispersed:
$$
\Delta p = \infty,
$$
so that Heisenberg relation is obviously satisfied. In the concentrated
state the localization in particle momentum can be small and the
(analog of) Heisenberg relation can be not satisfied
$$
\kappa = \frac{2}{\hbar} \Delta x \Delta p \ll 1.
$$
This does not mean breakdown of Heisenberg relation, because it is
a~mathematical fact, but concerning $\Delta p^{(QM)}$.

On the other hand, physical consequences of the short-time evolution
of~concentrated states contradict Heisenberg principle. The evolution
proceeds like an evolution with almost defined space and the (particle)
momentum localization, i.e. close to the DetQM-evolution.

This type of behavior needs two assumptions to be satisfied: \linebreak
the~starting state must be a concentrated state and the time interval of the
evolution must be short. Under these conditions Heisenberg principle
is broken in subquantum models.

(C) The preparation of the concentrated states proceeds by applying
(passing through) the iterated slits.

Passing through the slit $(a, b)$ of time $t$ is represented by
the~projection
$$
\psi (p, x; t) \mapsto \psi'(p, x; t) = \psi (p, x; t) . \chi_{a, b} (x).
$$
In the calculation below we approximate the projection by the
multiplication by a~"Gaussian" slit $\widetilde \chi_{a, b}$
$$
\chi_{a, b} (x) \approx \widetilde \chi (x_0, \Delta x; x) = c . \exp
\left\{ - \frac{1}{2} \frac{(x - x_0)^2}{\Delta x^2} \right\},
\tag{7.6}
$$
where the center $x_0$ and the dispersion $\Delta x$ are defined by
$$
x_0 = \frac{1}{2} (a + b), \ \Delta x = \frac{1}{2} (b - a)
$$
and the normalization constant $c$ is defined by the condition
$$
\int \widetilde \chi (x_0, \Delta x; \cdot) dx = \int \chi_{a, b}dx = b - a =
2 \Delta x.
$$

We shall denote the "passing the slit" projection by
$$
\psi (p, x; t) \mapsto \psi' (p, x; t) = \psi (p, x; t) . \widetilde
\chi (x_0, \Delta x; x).
$$

The "iterated slit" process consists in the following
$$
\spreadlines{-2pt}
\multline
\psi_1 (p^1, x^1; 0) \mapsto \psi'_1 (p^1, x^1; 0) = \psi_1 (p^1, x^1; 0) \,
\widetilde \chi (x_0^1, \Delta x^1; x^1) \mapsto \\
\mapsto \psi'_2 (p^2, x^2; T) =
\int K_T (p^1, x^1; p^2, x^2) \, \psi'_1 (p^1, x^1; 0) dp^1 dx^1 \mapsto \\
\mapsto \psi''_2 (p^2, x^2; T) = \psi'_2 (p^2, x^2; T) \,
\widetilde \chi (x_0^2, \Delta x^2; x^2).
\endmultline
\tag{7.7}
$$
This means that we apply the slit $\widetilde \chi (x_0^1, \Delta x^1;
\cdot)$ at time $t = 0$ and then the second slit $\widetilde \chi
(x_0^2, \Delta x^2; \cdot)$ at~time $t = T$. We shall show that if $\Delta
x^1$ and $\Delta x^2$ are sufficiently small and if the evolution time
$T$ is sufficiently small, $\beta T \ll 1$, then the resulting wave
function $\psi''_2$ describes the concentrated state.

This will be done in two steps. At the first step we shall show that the
short-time evolution $\psi'_2(\cdot, \cdot; T)$ contains the term
bounding together $x^2$ and $p^2$,
$$
|\psi'_2(p^2, x^2; T)| \leq const . \exp \left\{ - \frac{(x^2 - p^2
T/m)^2}{const (\Delta x^1)^2} \right\}. \tag{7.8}
$$
This comes from the fact that $p^1$ cannot differ much from $p^2$ and
then the dispersion of~$x^2 - p^2 T / m$ is also small. We shall show
this below.

The second (more simple) step combines together the term bounding $x^2$
and $p^2$ and the second slit term bounding $x^2$ to $x_0^2$. This
creates simply the term bounding $p^2 T / m$ to $x_0^2 - x_0^1$. This
creates the~completely concentrated state.
The corresponding calculations will be described below.

In the case when $\delta = \Delta x_1 = \Delta x_2$ (both slits are
of~the~same dimension) we obtain
$$
\kappa^2 = \frac{4}{\hbar^2} \delta^2 \Delta p_2^2 \cong 8 \frac{m^2
\delta^4}{T^2 \hbar^2} + \frac{4}{9} (\beta T)^4.
$$

(D) The short-time evolution of the concentrated state.

Let us consider the concentrated state $\psi_1(p, x)$ satisfying the
inequality (7.4) and let us assume that the evolution time $T$ satisfies
the relation $\beta T \ll 1$. Then the evolved state
$$
\psi_2 (p^2, x^2; T) = \int K_T (p^1, x^1; p^2, x^2) \psi_1 (p^1, x^1)
dp^1 dx^1
$$
satisfies the relation
$$
|\psi_2| \leq c . \exp \left\{- \frac{1}{2} \frac{1}{\Delta x_2^2} \left(
x^2 - x_0^1 - \frac{p_0}{m} T \right)^2 \right\} \tag{7.9}
$$
where the dispersion $\Delta x_2^2$ at the time $T$ is given by (the
calculation can be found below)
$$
\Delta x_2^2 = \Delta x_1^2 + \Delta p_1^2 \frac{T^2}{m^2} + \frac{1}{9}
(\beta T)^6 \frac{\hbar^2}{\Delta p_1^2 (\beta T)^2 + \Delta x_1^2 m^2
\beta^2}. \tag{7.10}
$$
This has to be confronted with the QM-evolution (assuming the relaxed
distribution of~$\psi_1$ in~$p^1$, or equivalently, that $\beta
\rightarrow \infty$) which gives the standard result
$$
\Delta x_2^{2 \ (QM)} = \Delta x_1^2 + \frac{\hbar^2}{4 \Delta x_1^2}
\cdot \frac{4 T^2}{m^2}. \tag{7.11}
$$

This QM-result is completely consistent with Heisenberg principle, by
which
$$
\Delta p_1^{2 \ (QM)} \approx \frac{\hbar^2}{4 \Delta x_1^2}.
$$
We see that in the both cases (7.10) and (7.11), the velocity of the growth
of $\Delta x_2^2$ is proportional to $\Delta p_1^2$, but this quantity
can be \linebreak
arbitrarily small in the subquantum models, while in the QM-case,
the quantity $\Delta p_1^{2 \ (QM)}$ is strictly bounded from below by
the Heisenberg principle. We have
$$
\Delta p_1^{2 \ (QM)} \geq \frac{\hbar^2}{4 \Delta x_1^2}
$$
in QM-case but it can happen
$$
\Delta p_1^2 \ll \frac{\hbar^2}{4 \Delta x_1^2}
$$
in the subquantum case. We shall call this situation {\it the
concentration effect in~subquantum models}. As a result we obtain
that
$$
\Delta x_2^2 \ll \Delta x_2^{2 \ (QM)} \tag{7.12}
$$
and this is the quantitative consequence of the concentration effect.

The corresponding gedanken experiment showing the observational
difference between QM and SubQM$_{RF}$ is the following. We let pass the
beam of~particles through iterated slits (making, e.g. $x_0 = 0$, $p_0 =
0$) and we shall observe the particle on the screen behind the~slits.
Assuming that the~slits are sufficiently narrow and that time intervals
of~the~evolution between slits and between the last slit and the screen
is sufficiently small, we obtain as an observational fact inequality
(7.12) expressing the clear difference between subquantum model and QM.
More details on this type of experiments can be found in the last
section.

It can be seen that this type of the effect is completely necessary
in~all subquantum models. The behavior on short time intervals is closer
to~the deterministic model DetQM and this implies both that the
concentration states are created by passing through iterated slits and
that the short-time evolution of~the~concentrated state will contradict
Heisenberg principle. Thus this type of gedanken experiment is
possible in all subquantum models. The order of quantities used in these
gedanken experiments depends on $\beta$ and the order of~$\beta$ is
hypothetical.

(E) In this last part we shall describe detailed calculations giving
formulas used above.

(i) Calculations will be done by using the following formulas. Let
$$
X, \ E_k, \ B_0 \in \Bbb R^2, \ k~= 1, \dots, K,
$$
be two-dimensional vectors,
$$
a_k, \ b_k, \ a_0 \in \Bbb C, \ k~= 1, \dots, K,
$$
be numbers satisfying $\roman{Im} (a_k) \geq 0$. Let $\otimes$ denote the tensor
product,
$$
\left( \matrix
       a\\
       b
       \endmatrix \right) \otimes
\left( \matrix
       c & d
       \endmatrix \right) =
\left( \matrix
       ac & ad\\
       bc & bd
       \endmatrix \right)
$$
and let $\bot$ denote the 90-degree rotation,
$$
\left( \matrix
       a\\
       b
       \endmatrix \right)^\bot =
\left( \matrix
       b\\
       -a
       \endmatrix \right).
$$
Then we have
$$
\multline
\int \exp \left\{ \frac{i}{2 \hbar} \sum_{k=1}^K a_k \left( X^\top e_k +
b_k \right)^2 + \frac{i}{\hbar} a_0 X^\top B_0 \right\} d^2 X = \\
= const. \exp \frac{i}{2 \hbar} \left\{ - \Delta^{-1} Y^{\top} \Big(
\sum_k a_k E_k^\bot \otimes {E_k^\bot}^\top \Big) Y + \sum_k a_k
b_k^2 \right\},
\endmultline
$$
where
$$
Y = \sum_{k=1}^K a_k b_k E_k + a_0 B_0, \ \Delta = det \left( \sum_k a_k E_k
\otimes E_k^\top \right).
$$

In the one-dimensional situation we have the special case
$$
\spreadlines{-2pt}
\multline
\int \exp \left\{ \frac{i}{2 \hbar} \sum a_k (x + b_k)^2 +
\frac{i}{\hbar} b_0 x \right\} d^1 x =\\
= const. \exp \frac{i}{2 \hbar} \left\{ - \Big( \sum_k a_k \Big)^{-1}
\left( \sum a_k b_k + b_0 \right)^2 + \sum_k a_k b_k^2 \right\}.
\endmultline
$$

(ii) We shall show that $\psi_2 (p^2, x^2; T)$ contains the
concentration of~$(x^2 - T m^{-1} p^2)^2$. We shall use the short-time
propagator described above and assume that $\beta T \ll 1$. Let the
dimensionless quantity $\sigma_1$,
$$
\frac{\beta m}{\hbar} \sigma_1 = \frac{1}{\Delta x_1^2},
$$
describe the extension of the first and second slit. We have to
calculate the following Gaussian integral
$$
\spreadlines{-2pt}
\multline
\int \exp \frac{i}{2 \hbar} \Big\{ \beta m i \sigma_1 x_1^2 +
\frac{1}{\beta m \beta T} (p_1 - p_2)^2 + \\
+ \beta m \omega_T \Big( x_1 - x_2 + \frac{T}{2m} (p_1 + p_2) \Big)^2 \Big\}
dp_1 dx_1,
\endmultline
$$
where
$$
\omega_T = \frac{12}{(\beta T)^3},
$$
and we use the simpler notation with lower indices, $x^1 \rightarrow
x_1$, $p^1 \rightarrow p_1$ etc.
At first we shall change the third term using the formula for $\omega_T$
to the form
$$
\frac{3}{\beta m \beta T} \left( \frac{2m}{T} x_1 - \frac{2m}{T} x_2 +
p_1 + p_2 \right)^2
$$
and using the change of variables
$$
p_1 \rightarrow p_1 - p_2 + \frac{2m}{T} x_2,
$$
we arrive at the integral
$$
\spreadlines{-2pt}
\multline
\int \exp \frac{i}{2 \hbar} \Big\{ \beta m i \sigma_1 x_1^2 +
\frac{1}{\beta m \beta T} \Big( p_1 -  2 p_2 + 2 \frac{m}{T} x_2 \Big)^2 + \\
+ 3 \frac{\beta m}{\beta T} \Big( \frac{2m}{T} x_1 + p_1 \Big)^2 \Big\}
dp_1 dx_1.
\endmultline
$$

Now we shall use the formula from the preceding step (i) with $K = 3$. We
have
$$
\align
&a_1 = \beta m i \sigma_1,\ a_2 = \frac{1}{\beta m \beta T},\
 a_3 = \frac{3}{\beta m \beta T},\\
&b_1 = 0,\ b_2 = 2 \left( - p_2 + \frac{m}{T} x_2 \right),\ b_3 = 0, \\
&E_1 = \left( {\matrix
       0\\
       1
      \endmatrix} \right),\
E_2 = \left( {\matrix
       1\\
       0
      \endmatrix} \right),\ \
E_3 = \left( {\matrix
        1\\
        2 m / T
       \endmatrix} \right).
\endalign
$$
Then we have
$$
\align
Y = \sum a_k b_k E_k &= \frac{2}{(\beta T)^2} \left( \matrix
                                                     x_2 - T p_2 / m \\
                                                     0
                                                    \endmatrix \right),\\
\sum a_k b_k^2 &= \frac{4 \beta m}{(\beta T)^3} \left( x_2 - \frac{T}{m}
p_2 \right)^2,\\
\sum a_k E_k \otimes E_k^\top \!&=\! \frac{1}{\beta m (\beta T)^3}
\left( \! \matrix
         4 (\beta T)^2 &6 \beta m \beta T\\
         6 \beta m \beta T &12 (\beta m)^2 \!+\! i \sigma_1 (\beta m)^2
         (\beta m)^3
       \endmatrix \! \right), \\
\Delta^{-1}\! &=\! \frac{3}{4 (\beta m)^2 (\beta T)^2 (9 \!+\! \sigma_1^2 (\beta
T)^6)} \!-\! \frac{i \sigma_1 \beta T}{4 (\beta m)^2 (9 \!+\! \sigma_1^2 (\beta
T)^6)}.
\endalign
$$
For the inverse matrix we obtain the formula
$$
\multline
\left( \sum a_k E_k \otimes E_k^\top \right)^{-1} = \\
= \frac{6 T}{4 m (9 + \sigma_1^2 (\beta T)^6)} \cdot
\left( \matrix
        (\beta m)^2 (6 + \sigma_1^2 (\beta T)^6 \cdot \frac{1}{6})
        &- 3 \beta m \beta T\\
        - 3 \beta m \beta T &2 (\beta T)^2
       \endmatrix \right) - \\
- \frac{i \sigma_1 (\beta T)^4}{4 \beta m (9 + \sigma_1^2 (\beta T)^6)}
\left( \matrix
        9 (\beta m)^2 &- 6 \beta m \beta T\\
        - 6 \beta m \beta T &4 (\beta T)^2
       \endmatrix \right).
\endmultline
$$

To obtain the resulting concentration we need only the imaginary part
of~this matrix:
$$
\spreadlines{-2pt}
\multline
\roman{Im} \left( \sum a_k E_k \otimes E_k^\top \right)^{-1} = \\
= \frac{- i \sigma_1 (\beta T)^2}{4 \beta m (9 + \sigma_1^2 (\beta T)^6)}
\left( \matrix
        3 \beta m\\
        - 2 \beta T
       \endmatrix \right) \otimes
\left( \matrix
        3 \beta m, &- 2 \beta T
       \endmatrix \right).
\endmultline
$$
Then we obtain
$$
Y^\top \roman{Im} \left( \sum a_k E_k \otimes E_k^\top \right)^{-1} Y =
\frac{- i \sigma_1 \beta m}{1 + \frac{1}{9} \sigma_1^2 (\beta T)^6}
\left( x_2 - \frac{T}{m} p_2 \right)^2
$$
and the concentration term
$$
\exp \left\{ - \frac{1}{2 \hbar} \cdot \frac{\sigma_1 \beta m}{1 + \frac{1}{9}
\sigma_1^2 (\beta T)^6} \left( x_2 - \frac{T}{m} p_2 \right)^2 \right\}.
$$

Applying the term $- i / 2 \hbar$ and the expression of $\sigma_1$
in~terms of~$\Delta x_1^2$ we obtain the concentration term
$$
\exp \left\{ - \frac{1}{2} \frac{1}{\widetilde{\Delta x_2}^2} \left( x_2 -
\frac{T}{m} p_2 \right)^2 \right\}
$$
with
$$
\widetilde{\Delta x_2}^2 = \Delta x_1^2 + \frac{\hbar^2}{9 \Delta x_1^2
(\beta m)^2} (\beta T)^6.
$$
For small $\beta T \ll 1$ we obtain the concentration of $(x_2 - T p_2
/ m)^2$ of~the~order of~$\Delta x_1^2$.

(iii) At this moment, applying the second slit at the time $T$ to the
wave function $\psi_2$, we obtain
$$
| \psi'_2 (p_2, x_2; T)| = | \psi_2 (p_2, x_2; T)| . const . \exp
\left\{ - \frac{1}{2} \frac{1}{\Delta x_2^2} x_2^2 \right\}.
$$

From part (ii) we know that
$$
| \psi'_2 (p_2, x_2; T)| \leq const . \exp - \frac{1}{2} \left[
\frac{1}{\widetilde{\Delta x_2}^2} \left( x_2 - \frac{T}{m} p_2 \right)^2
+ \frac{1}{\Delta x_2^2} x_2^2 \right]
$$
with
$$
\widetilde{\Delta x_2}^2 = \Delta x_1^2 + \frac{\hbar^2 (\beta T)^6}{9
\Delta x_1^2 (\beta m)^2}.
$$
From the inequality
$$
a (x_2 - \xi)^2 + b x_2^2 \geq \frac{ab}{a + b} \xi^2
$$
true for $a + b > 0$, we obtain
$$
\frac{1}{\widetilde{\Delta x_2}^2} \left( x_2 - \frac{T}{m} p_2 \right)^2
+ \frac{1}{\Delta x_2^2} x_2^2 \geq \frac{1}{\Delta p_2^2} p_2^2
$$
where
$$
\Delta p_2^2 = \left( \widetilde{\Delta x_2}^2 + \Delta x_2^2 \right)
\frac{m^2}{T^2}.
$$
Using the formula for $\widetilde{\Delta x_2}^2$ we arrive at
$$
\Delta p_2^2 \frac{T^2}{m^2} = \Delta x_1^2 + \Delta x_2^2 +
\frac{\hbar^2 (\beta T)^6}{9 \Delta x_1^2 (\beta m)^2}
$$
and
$$
|\psi'_2 (p_2, x_2; T)| \leq const . \exp - \frac{1}{2} \left\{
\frac{1}{\Delta x_2^2} x_2^2 + \frac{1}{\Delta p_2^2} p_2^2 \right\}.
$$

For $\Delta x_1 = \Delta x_2$ we obtain the~degree of~concentration
$$
\kappa^2 = \frac{4}{\hbar^2} \Delta x_2^2 \Delta p_2^2 = 8
\frac{m^2}{T^2 \hbar^2} \cdot \Delta x_2^4 + \frac{4}{9} (\beta T)^4.
$$
In the "equilibrated" situation, where first and second terms are
of~the~same order, we have
$$
\Delta x_1 = \Delta x_1 \cong \frac{1}{2} \beta T \left( \frac{T
\hbar}{m} \right)^{1/2}
$$
and
$$
\kappa \cong (\beta T)^2.
$$


In this way the concentrated states can be prepared.

(iv) Let us look for a concentration state which is centered around
$x_{20}$ and $p_{20}$, so that the inequality
$$
|\psi'(p_2, x_2 ;T)| \leq const . \exp - \frac{1}{2} \left\{ \frac{(x_2
- x_{20})^2}{\Delta
x_2^2} + \frac{(p_2 - p_{20})^2}{\Delta p_2^2} \right\}
$$
is fulfilled.

Of course, we suggest that the first slit has to be centered around
$$
x_{10} := x_{20} - \frac{p_{20}}{m} T.
$$
Then the relevant integral is the following
$$
\multline
\int \exp \frac{i}{2 \hbar} \Big\{ i \frac{(x_1 - x_{10})^2}{\Delta
x_1^2} + i \frac{(x_2 - x_{20})^2}{\Delta x_2^2} + \frac{1}{\beta m
\beta T} (p_1 - p_2)^2 + \\
+ \beta m \omega_T \Big( x_1 - x_2 + \frac{T}{2m} (p_1 + p_2) \Big)^2
\Big\} dp_1 dx_1.
\endmultline
$$
We shall make the substitution
$$
\alignat2
&x_1 = \overline x_1 + x_{20} - \frac{p_{20}}{m} T, \ \ && x_2 = \overline x_2 +
x_{20},\\
&p_1 = \overline p_1 + p_{20}, \ && p_2 = \overline p_2 + p_{20}
\endalignat
$$
and obtain after the change $dp_1 dx_1 = d \overline p_1 d \overline x_1$
$$
\spreadlines{-2pt}
\multline
\int \exp \frac{i}{2 \hbar} \Big\{ i \frac{\overline x_1^2}{\Delta x_1^2}
+ i \frac{\overline x_2^2}{\Delta x_2^2} + \frac{1}{\beta m \beta T}
(\overline p_1 - \overline p_2)^2 + \\
+ \beta m \omega_T \Big( \overline x_1 - \overline x_2 + \frac{T}{2m}
(\overline p_1 + \overline p_2) \Big)^2 \Big\} d \overline p_1 d \overline x_1.
\endmultline
$$
By the result of (iii) we obtain
$$
|\psi'_2(\overline p_2, \overline x_2; T)| \leq const . \exp - \frac{1}{2} \left\{
\frac{\overline x_2^2}{\Delta x_2^2} + \frac{\overline p_2^2}{\Delta p_2^2} \right\}
$$
and after the change $\overline x_2 \rightarrow x_2 - x_{20}$, $\overline p_2
\rightarrow p_2 - p_{20}$ we arrive at the formula we were looking for.

(v) Now we shall start with the concentrated state
$$
|\psi_1 (p_1, x_1; 0)| \leq const . \exp - \frac{1}{2} \left\{
\frac{(x_1 - x_{10})^2}{\Delta x_1^2} + \frac{(p_1 - p_{10})^2}{\Delta p_1^2}
\right\}
$$
and it will be evaluated to
$$
\psi_2 (p_2, x_2; T) \leq \int K_T (p_1, x_1; p_2, x_2) \psi_1 (p_1,
x_1; 0) dp_1 dx_1.
$$
We are interested mainly in the localization of $\psi_2$ in $x_2$ in the
form of~the~interpretation from Section 5.
$$
\left| \int \psi_2 (p_2, x_2; T) dp_2 \right| \leq const . \exp \left\{ - \frac{1}{2}
\frac{(x_2 - x_{20})^2}{\Delta x_2^2} \right\}.
$$

It is sufficient to make an integration on $dp_2$ inside the propagator
and to obtain the reduced propagator
$$
\widetilde{K_T} (p_1, x_1; x_2) := \int K_T (p_1, x_1; p_2, x_2) dp.
$$
We shall obtain
$$
\widetilde{K_T} (p_1, x_1; x_2) = \exp \left\{ \frac{i}{2 \hbar} \beta m
\widetilde{\omega_T} \left( x_2 - x_1 - \frac{T}{m} p_1 \right)^2
\right\},
$$
where
$$
\widetilde{\omega_T} = \frac{3}{(\beta T)^3}.
$$
We have to calculate the integral
$$
\spreadlines{-3pt}
\multline
\int \exp \frac{i}{2 \hbar} \Big\{ \frac{1}{\beta m \beta T} (p_2 -
p_1)^2 + \\
+ \beta m \frac{12}{(\beta T)^3} \Big( x_2 - x_1 - \frac{T}{2m}
(p_1 + p_2) \Big)^2 \Big\} dp_2.
\endmultline
$$
The second term may be rewritten as
$$
\frac{3}{\beta m \beta T} \left( p_2 + p_1 - (x_2 - x_1) \frac{2m}{T}
\right)^2.
$$
Using the last formula from (i) for $K = 2$ we obtain the formula
for~$\widetilde{K_T}$.

(vi) In the calculation of the short-time evolution of the concentrated
state we shall use the reduced propagator $\widetilde{K_T}$. We have
to~calculate the following integral (where $\widetilde{\omega_T} = 3 (\beta
T)^{-3}$)
$$
\spreadlines{-3pt}
\multline
\int \exp \frac{i}{2 \hbar} \Big\{ i \beta m \sigma_1 (x_1 - x_{10})^2 + \\
+ i \frac{\rho_1}{\beta m} (p_1 - p_{10})^2 + \beta m \widetilde{\omega_T}
\Big( x_1 - x_2 + \frac{T}{m} p_1 \Big)^2 \Big\} dx_1 dp_1.
\endmultline
$$
We make substitutions
$$
\align
p_1 &\rightarrow p_1 + p_{10}, \\
x_1 &\rightarrow x_1 + x_{10}, \\
\overline x_2 &= x_2 - x_{10} - \frac{T}{m} p_{10}
\endalign
$$
and then we obtain the integral
$$
\int \exp \frac{i}{2 \hbar} \left\{ i \beta m \sigma_1 x_1^2
+ i \frac{\rho_1}{\beta m} p_1^2 + \beta m \widetilde{\omega_T}
\left( x_1 + \frac{T}{m} p_1 - \overline x_2 \right)^2 \right\} dx_1 dp_1.
$$
Now we shall apply the formula from (i) with $K = 3$ and
$$
\align
&a_1 = i \beta m \sigma_1, \ a_2 = i \frac{\rho_1}{\beta m},\ a_3 =
\beta m \widetilde{\omega_T}, \\
&b_1 = b_2 = 0, \ b_3 = - \overline x_2, \\
&E_1 = \left( {\matrix
                0\\
                1
               \endmatrix} \right), \
E_2 = \left( {\matrix
              1\\
              0
             \endmatrix} \right), \
E_3 = \left( {\matrix
              T/m\\
              1
             \endmatrix} \right).
\endalign
$$
Then we obtain
$$
\multline
A^{-1} := \left( \sum a_k E_k \otimes E_k^\top \right)^{-1} = \\
= \Delta^{-1} \frac{1}{\beta m} \left[ \widetilde{\omega_T} \left( \matrix
                                                      \beta m \\
                                                      - \beta T
                                                    \endmatrix \right)
\otimes \left( \matrix
                \beta m &- \beta T
               \endmatrix \right)
+ i \left( \matrix
            \sigma_1 (\beta m)^2 &0\\
            0 &p_1
           \endmatrix \right) \right],
\endmultline
$$
and where
$$
\Delta^{-1} = \frac{- \rho_1 \sigma_1 - i \widetilde{\omega_T} (\sigma_1
\beta^2 T^2 + \rho_1)}{(\rho_1 \sigma_1)^2  + {\widetilde{\omega_T}}^2
(\sigma_1 \beta^2 T^2 + \rho_1)^2}.
$$
Using
$$
Y = \sum a_k b_k E_k = - \widetilde{\omega_T} \overline x_2 \left(
\matrix
 \beta T\\
 \beta m
\endmatrix \right)
$$
we obtain
$$
Y^\top A^{-1} Y = \Delta^{-1} i \beta m {\widetilde{\omega_T}}^2 (\sigma_1
\beta^2 T^2 + \rho_1) \overline x_2^2.
$$

We are interested only in the localization term so that only
\linebreak
the~imaginary part contributes (the term $\sum a_k b_k^2$ contributes
\linebreak
to~the~real part),
$$
\roman{Im} (Y^\top A^{-1} Y) = - \overline x_2^2 \beta m
\frac{\widetilde{\omega_T}^2 (\sigma_1 \beta ^2 T^2 + \rho_1) \sigma_1
\rho_1}{(\sigma_1 \rho_1)^2 + \widetilde{\omega_T}^2 (\sigma_1 \beta ^2 T^2
+ \rho_1)^2}.
$$
From the equality
$$
\exp \left\{ - \frac{1}{2} \frac{\overline x_2^2}{\Delta x_2^2} \right\} =
\exp \left\{ \frac{1}{2 \hbar} \roman{Im} (Y^\top A^{-1} Y) \right\}
$$
and from equations
$$
\align
\frac{1}{\sigma_1} &= \frac{\beta m}{\hbar} \Delta x_1^2, \\
\frac{1}{\rho_1} &= \frac{1}{\hbar \beta m} \Delta p_1^2
\endalign
$$
we obtain the final formula for the dispersion $\Delta x_2^2$
of~the~wave packet at~the~time $T$
$$
\Delta x_2^2 = \Delta x_1^2 + \frac{T^2}{m^2} \Delta p_1^2 + \frac{1}{9}
\frac{T^2}{m^2} (\beta T)^4 \frac{\hbar^2}{\Delta x_1^2 + \frac{T^2}{m^2}
\Delta p_1^2}.
$$

(vii) The analogical formula for the dispersion $\Delta x_2^2$ in QM is
standard. The integral to be calculated is
$$
\int \exp \frac{i}{2 \hbar} \left\{ i \hbar \frac{(x_1 - x_{10})^2}{\Delta x_1^2}
+ \frac{m}{T} (x_1 - x_2)^2 \right\} dx_1.
$$
Making the substitution $x_1 \rightarrow x_1 + x_{10}$, $\overline x_2
= x_2 - x_{10}$ we obtain the~integral
$$
\int \exp \frac{i}{2 \hbar} \left\{ \frac{i \hbar}{\Delta x_1^2} x_1^2
+ \frac{m}{T} (x_1 - \overline x_2)^2 \right\} dx_1.
$$
Using the last formula from (i) we obtain that the real part
of~the~resulting Gaussian is
$$
\exp \left\{ - \frac{1}{2} \frac{\overline x_2^2}{\Delta x_2^2} \right\}
$$
where
$$
\Delta x_2^2 = \Delta x_1^2 + \frac{\hbar^2}{\Delta x_1^2} \cdot
\frac{T^2}{m^2}.
$$

If we introduce the conjugated quantity
$$
\Delta {p_1^2}^{(QM)} = \frac{\hbar^2}{4 \Delta x_1^2}
$$
then we have an analogical formula
$$
\Delta x_2^2 = \Delta x_1^2 + \frac{4 T^2}{m^2} \Delta {p_1^2}^{(QM)}.
$$
This quantity satisfies, of course, the Heisenberg relation.


\newpage
\head
8. The correlated random force model SubQM$_{CRF}$ \\
and the correlation effect
\endhead

The starting point of~subquantum model was the idea of~the~SLO-vacuum --
the~medium composed from space-like objects. The random force served as
a~model of~the~interaction of~the~system with such a~medium. We have
supposed that the random forces $F (t, \vec x_1)$ and~$F (t, \vec x_2)$,
$\vec x_1 \not= \vec x_2$, representing interaction with
the~SLO-vacuum, are stochastically independent.

The opposite hypothesis, that these forces are not completely
independent, is also possible. Let us consider the model
of~the~space-like objects with the~zero (space-like) velocity
$$
t = f^\alpha (\vec x) \equiv t_0^\alpha,\  \alpha \in \Bbb Z.
$$
One can then think on idea that the random force is the same
at~different places in the space and that it depends only on~the~time
$$
F (t, \vec x) \equiv F_0 (t)
$$
and then the forces $F_i (t) = F (t, \vec x_i)$ and $F_j (t) = F (t,
\vec x_j)$ , $i \not= j$, are equal.

The completely opposite assumption that the random forces $F_i (t)$ and
$F_j (t)$, $\forall i, j$, are the same is too strong. We shall suppose that the random
forces will contain the part $G_0$ which is the same for~all particles
and the~part $G_i$ which is different for different particles. The
hypotheses will be the following; they substitute hypotheses (i)-(iv)
from Section 3:

(i) There exists a~random force $F_i (t)$ acting on~the i-th degree
of~freedom, $i = 1, \dots, n$.

(ii) Forces $F_i (t)$ can be expressed as
$$
F_i (t) = G_0 (t) + G_i (t), \ i = 1, \dots, n,
$$
where forces $G_0 (t)$, $G_1 (t)$, \dots, $G_n (t)$ are statistically
independent.

(iii) There is an~amplitude distribution of the random forces given by
$$
\align
\Cal A_{t^1, t^2} [G_0] &= \exp \left\{ \frac{i}{\hbar} \frac{a_0}{2}
\int_{t^1}^{t^2} G_0^2 (t) dt \right\} \prod_t dG_0 (t),\\
\Cal A_{t^1, t^2} [G_j] &= \exp \left\{ \frac{i}{\hbar} \frac{a_1}{2}
\int_{t^1}^{t^2} G_j^2 (t) dt \right\} \prod_t dG_j (t), \ j = 1, \dots, n.
\endalign
$$

(iv) The system with $n$ degrees of~freedom is described by DetQM with
a~given random force.

Forces $F_i (t)$ and $F_j (t)$, $i \not= j$, are correlated, because they
both contain the common part $G_0 (t)$, while other parts $G_i (t)$
and~$G_j (t)$ are independent. We shall proceed in the following steps.
\roster
\item"(A)" The Feynman integral,
\item"(B)" the propagator,
\item"(C)" definition and preparation of the correlated states,
\item"(D)" evolution of the correlated states.
\endroster

(A) The Feynman integral for the transition amplitude in the
SubQM$_{CRF}$
model is
$$
\spreadlines{-2pt}
\multline
\Cal A~= \int_{(BC)} \exp \left\{ \frac{i}{\hbar} \Cal A_{t^1, t^2}
[x_1, \dots, x_n] \right\} \prod_{i, t} \delta \big( m_i \ddot x_i (t) - G_0 (t) -
G_i (t) \big) .\\
. \exp \frac{i}{2 \hbar} \left\{ \int_{t^1}^{t^2} \Big( a_0 G_0^2 (t) +
a_1 \sum_{i = 1}^n G_i^2 (t) \Big) dt \right\} . \\
. \prod_t dG_0 (t) \prod_{t, i} dG_i (t) \prod_{t, i} dx_i (t).
\endmultline
$$
The boundary conditions are standard: $x_i (x^s) = x_i^s$, $m_i \dot x_i
(t^s) = p_i^s$, $s = 1, 2$, $i = 1, \dots, n$. For simplicity we shall
suppose that
$$
\align
m_i &= m \ (\forall i), \\
\Cal A~[x_i] &= \int_{t^1}^{t^2} \Big( \frac{m}{2} \sum_{i=1}^n
\dot x_i^2 (t) - V~(x_i(t)) \Big) dt, \\
n &= 3n_0,
\endalign
$$
i.e. that we have an interacting system of~$n_0$ particles subjected
to~correlated random forces and all particles have the same mass $m$.

At first we shall make the integration with respect to $\prod dG_i (t)$.
The $\delta$-functions imply that $G_i = m \ddot x_i - G_0$, so that we
shall obtain the integral
$$
\multline
\int_{(BC)} \exp \left\{ \frac{i}{\hbar} \int \frac{m}{2} \sum \dot
x_i^2 - V~(x_i) dt \right\} . \\
. \exp \frac{i}{2 \hbar} \left\{ \int \Big( a_0 G_0^2 +
a_1 \sum (m \ddot x_i - G_0)^2 \Big) dt \right\} .
\prod_t dG_0 (t) \prod_{t, i} dx_i (t)
\endmultline
$$
and this gives
$$
\multline
\!\!\!\!\!\!\! \int_{(BC)} \!\!\!\! \exp \frac{i}{2 \hbar} \left\{ \int
\Big( m \sum \! \dot x_i^2 \!-\! 2 V~(x_i) \!+\! a_1 m^2 \sum \! \ddot x_i^2
\!+\! (a_0 \!+\! n a_1) G_0^2 \Big) dt \right\} .\\
. \exp \left\{ - \frac{i}{\hbar} \int a_1 m G_0 \sum \ddot x_i dt
\right\} \prod_t dG_0 (t) \prod_{t, i} dx_i (t).
\endmultline
$$
Integrating with respect to $\prod dG_0 (t)$ we obtain finally
$$
\multline
\Cal A~= \int_{(BC)} \exp \Big\{ \frac{i}{2 \hbar} \int \Big( m \sum
\dot x_i^2 - 2 V~(x_i) +\\
+ a_1 m^2 \sum \ddot x_i^2 + a_2 m^2 \big( \sum \ddot x_i \big)^2 \Big) dt \Big\}
\prod_{t, i} dx_i (t),
\endmultline
$$
where
$$
a_2 = - \frac{a_1^2}{a_0 + n a_1}.
$$

We see that the collective term with $a_2$ is a~new feature of this model.
If $a_2 = 0$ and $a_1 = a$, this model is the same as SubQM$_{RF}$. The term
$(\sum \ddot x_i)^2$ creates certain interaction among particles .

(B) In the calculation of~the~propagator we shall assume that
the~interaction term is zero,
$$
V~\equiv 0.
$$
Our way to~diagonalize the new term requires to do the~orthogonal
transformation
$$
x_i (t) = \sum_{j=1}^n R_{ij} y_j (t)
$$
such that
$$
R_{in} = (n)^{- 1/2} \text{ for } i = 1, \dots, n.
$$
Columns of~the~orthogonal matrix $R$ compose a~basis of~$\Bbb R^n$ and
they are, for example,
$$
\align
R_{ij} &= (i^2 + i)^{-1/2} \text{ for } j \leq i \leq n - 1, \\
R_{nj} &= n^{-1/2} \text{ for } j \leq n, \\
R_{j, j+1} &= - j (j^2 + j)^{-1/2} \text{ for } j \leq n - 1, \\
R_{ij} &= 0 \text{ for } j \geq i + 2.
\endalign
$$

The inverse transformation is
$$
y_k (t) = \sum_i R_{ik} x_i (t).
$$
So that $R^\top R = 1$. We shall introduce the corresponding
boundary conditions for $y_j$`s
$$
\roman{(BC)_Y: } \quad y_j (t^s) = y_j^s, \
m \dot y_j (t^s) = q_j^s, \
Y_j^s = \left( \matrix
                q_j^s \\
                y_j^s
               \endmatrix \right), \
s~= 1, 2,
$$
where
$$
y_j^s = \sum R_{ij} x_i^s, \
q_j^s = \sum R_{ij} p_i^s, \
Y_j^s = \sum R_{ij} X_i^s, \
s~= 1, 2.
$$
From orthogonality of~$R$ we obtain
$$
\align
\sum \dot x_i^2 (t) &= \sum \dot y_j^2 (t), \\
\sum \ddot x_i^2 (t) &= \sum \ddot y_j^2 (t), \\
\Big( \sum \ddot x_i \Big)^2 &= n \ddot y_n^2.
\endalign
$$
We shall make the orthogonal change of~variables in~the~Feynman integral
and we obtain
$$
\spreadlines{-2pt}
\multline
\Cal A~= \left\{ \prod_{i = 1}^{n - 1} \int_{(BC)_Y} \exp \left[
\frac{i}{2 \hbar} \int m \dot y_i^2 + a_1 m^2 \ddot y_i^2 dt \right]
\prod_t dy_i (t) \right\} . \\
. \int_{(BC)_Y} \exp \left[ \frac{i}{2 \hbar} \int m \dot y_n^2 + a_3 m^2
\ddot y_n^2 dt \right] \prod_t dy_n (t),
\endmultline
$$
where
$$
a_3 = a_1 + a_2 n = \frac{a_0}{n + a_0 / a_1} \approx \frac{a_0}{n}.
$$
Here we assume that
$$
\tau_0 \ll \tau_1, \
\tau_0 = (a_0 m)^{1/2}, \
\tau_1 = (a_1 m)^{1/2},
$$
i.e. that the relaxation time of~the~$G_0$ is shorter than
the~relaxation time of~the~$G_i$-forces. This means that for~$T$,
$\tau_0 \ll T \ll \tau_1$, the~$G_0$-process is already relaxed but
the~$G_i$-processes are not relaxed. Equivalently, we have $a_0 \ll a_1$
and thus also
$$
a_3 \approx n^{-1} a_0 \ll a_1.
$$
We shall introduce
$$
\tau_3 = (a_3 m)^{1/2}, \
\beta_3 = 1 / \tau_3, \
\beta_1 = 1 / \tau_1.
$$

The last Feynman integrals may be simply calculated if we shall use
formula (6.4) using matrices $Q_{inT}$, $Q_{trT}$ and $Q_{outT}$,
where also their parametrical dependence on~$\beta$ is denoted by
$Q_{inT}^{\beta_1}$ etc.

The resulting propagator is $K_T = N_T \exp i \hbar^{-1} S_T$, where
$$
\multline
S_T = \sum_{i = 1}^{n - 1} \left( \frac{1}{2} {Y_i^1}^\top Q_{inT}^{\beta_1}
Y_i^1 + {Y_i^1}^\top Q_{trT}^{\beta_1} Y_i^2 + \frac{1}{2} {Y_i^2}^\top
Q_{outT} Y_i^2 \right) + \\
+ \frac{1}{2} {Y_n^1}^\top Q_{inT}^{\beta_3} Y_n^1 + {Y_n^1}^\top
Q_{trT}^{\beta_3} Y_n^2 + \frac{1}{2} {Y_n^2}^\top Q_{outT} Y_n^2.
\endmultline
$$
To transform this quantity into $X$-variables we introduce
$$
\align
\overline X^s &= \frac{1}{n} \sum_{i=1}^n X_i^s, \ s~= 1,2,\\
\Delta X_i^s &= X_i^s - \overline X^s, \ s~= 1,2, \ i = 1, \dots, n.
\endalign
$$
In this way we obtain
$$
Y_n^s = n^{- 1/2} \sum x_j^s = n^{1/2} \overline X^s.
$$
Using the formula of~the~type ($Q$ is any $2 \times 2$ matrix)
$$
\sum_{i=1}^n {X_i^s}^\top Q X_i^r = \Delta {X_i^s}^\top Q \Delta X_i^r +
n {\overline X^s}^\top Q \overline X^r, \ s, r = 1, 2,
$$
and
$$
\sum_{i=1}^n {Y_i^s}^\top Q Y_i^r = \sum_{j=1}^n {X_j^s}^\top Q X_j^r,
\ s, r = 1, 2.
$$
We obtain that
$$
\sum_{i=1}^n {Y_i^s}^\top Q Y_i^r = \sum_{j=1}^n \Delta {X_j^s}^\top Q
\Delta X_j^r,\ s, r = 1, 2.
$$
As a result we obtain the formula
$$
\multline
\!\!\!\!\!\!\! S_T = \sum_{i=1}^n \left( \frac{1}{2} \Delta {X_i^1}^\top
Q_{inT}^{\beta_1} \Delta X_i^1 + \Delta {X_i^1}^\top Q_{trT}^{\beta_1}
X_i^2 + \frac{1}{2} \Delta {X_i^2}^\top Q_{outT}^{\beta_1} X_i^2 \right) +\\
+ n \left( \frac{1}{2} {\overline X_i^1}^\top Q_{inT}^{\beta_3} \overline X^1 +
{\overline X^1}^\top Q_{trT}^{\beta_3} \overline X^2 + \frac{1}{2} {\overline X^2}^\top
Q_{outT}^{\beta_3} \overline X^2 \right).
\endmultline
$$

To obtain the final form of~the~propagator we shall use formulas
of~the~type
$$
\sum_{i=1}^{n-1} (y_i^s)^2 = \sum_{i=1}^n (\Delta x_i^s)^2, \
\sum_{i=1}^{n-1} (p_i^s)^2 = \sum_{i=1}^n (\Delta p_i^s)^2, \
(y_n^s)^2 = n (\bar x^s)^2, \dots
$$
In this way we obtain
$$
\multline
S_T = \sum_{i=1}^n \frac{1}{2 \beta_1 m} \left( \frac{(\Delta p_i^1)^2 +
(\Delta p_i^2)^2}{\tanh \beta_1 T} - \frac{2 \Delta p_i^1 \Delta p_i^2}
{\sinh \beta_1 T} \right) + \\
+ \frac{n}{2 \beta_3 m} \left( \frac{(\bar p^1)^2 + (\bar p^2)^2}{\tanh
\beta_3 T} - \frac{2 \bar p^1 \bar p^2}{\sinh \beta_3 T} \right) + \\
+ \frac{m}{2T} \Big\{ \frac{\sum_{i=1}^n \left( \Delta x_i^2 - \Delta
x_i^1 - (\beta_1 m)^{-1} (\Delta p_i^1 + \Delta p_i^2) \tanh (\beta_1
T/2) \right)^2}{1 - (\beta_1 T/2)^{-1} \tanh (\beta_1 T/2)} -\\
- \frac{n \left( \bar x^2 - \bar x^1 - (\beta_3 m)^{-1} (\bar p^1 + \bar p^2)
\tanh (\beta_3 T/2) \right)^2}{1 - (\beta_3 T/2)^{-1} \tanh (\beta_3 T/2)}
\Big\}.
\endmultline
$$

The result is that the relative positions and relative momenta evolve
with the relaxation constant $\beta_1$ while the mean value of~po\-sitions
and~the~mean value of~momenta evolve with relaxation \linebreak
constant $\beta_3 \gg \beta_1$. This phenomenon creates certain time interval during which
mean values are already relaxed (i.e. long-time case) and
relative \-values are still not relaxed (i.e. short-time case) -- such $T$ that
$\beta_3 T \gg 1$ and~$\beta_1 T \ll 1$. For such times the correlation
effect happens.

(C) The correlated state is the $n$-particle state satisfying
the~inequality
$$
|\psi (p_i, x_i)| \leq c . \exp \left\{ - \frac{1}{2} \frac{\sum (\Delta
x_i)^2}{\Delta x^2} - \frac{1}{2} \frac{\sum (\Delta p_i)^2}{\Delta p^2}
\right\}
$$
for some positive constants $\Delta x$, $\Delta p$, $c$, where
$$
\Delta p \cdot \Delta x \ll \hbar.
$$
This means that in the correlated state the~relative positions
and~re\-lative momenta are concentrated in the sense of~the~preceding
section.

Preparation of the correlated state is similar to~preparation
of~concentrated states -- particles pass through repeated slits.
Assuming two slits preparation, one has to introduce into the original
Feynman integral two Gaussians representing the process of~passing
through slits,
$$
\exp \left\{ - \frac{1}{2} \frac{\sum_{i=1}^n (x_i^1)^2}{\Delta x^2}
- \frac{1}{2} \frac{\sum_{i=1}^n (x_i^2)^2}{\Delta x^2} \right\}.
$$

Now we shall make transformation of variables to~new variables $y$. From
the~term describing the~effect of~slits we obtain
$$
\exp \left\{ - \frac{1}{2} \frac{\sum_{i=1}^n (y_i^1)^2}{\Delta x^2}
- \frac{\sum_{i=1}^n (y_i^2)^2}{\Delta x^2} \right\}.
$$
Now, using results of~the~preceding section (part (C)) we obtain,
assuming $\beta_1 T \ll 1$, that variables $y_i^2$ and $q_i^2$, $i = 1,
\dots, n-1$, will be concentrated, because they evolve with
the~relaxation constant $\beta_1$. Making then the inverse
transformation $y, q \rightarrow x, p$, we obtain from
$$
|\psi'_2| \leq \exp \left\{ - \frac{1}{2} \sum_{i=1}^{n-1} \frac{(y_i^2)^2}
{\Delta x^2} + \frac{(q_i^2)^2}{\Delta p^2} \right\}
$$
the inequality
$$
|\psi'_2| \leq \exp \left\{ - \frac{1}{2} \sum_{i=1}^n \frac{(\Delta x_i^2)^2}
{\Delta x^2} + \frac{(\Delta p_i^2)^2}{\Delta p^2} \right\}.
$$
Dependence of $\psi'_2$ on variables $y_n$, $q_n$ or~$\bar x$, $\bar p$
need not be concentrated and in fact, assuming $\beta_3 T \gg 1$, it
will be relaxed.

(D) The evolution of the correlated state constructed above can be
analyzed in~terms of~variables $y_i$, $q_i$. The correlated state
$\psi_1$ expressed in variables $\hat{y_i}$ and~$\hat{q_i}$ satisfies
($\Delta x \cdot \Delta p \ll \hbar$)
$$
\big| \psi_1 (y_1,\! \dots \! , y_n, q_1,\!  \dots \! , q_n) \big| \! \leq \!
c . \exp \left\{ \!- \frac{1}{2 \Delta x_1^2} \sum_{i=1}^{n-1} (y_i^1)^2
\!-\! \frac{1}{2 \Delta p_1^2} \sum_{i=1}^{n-1} (q_i^1)^2 \! \right\}.
$$
If the evolution time $T = t^2 - t^1$ satisfies
$$
\beta_1 T \ll 1 \ll \beta_3 T,
$$
then the evolution in variables $y_i$, $q_i$, $i \leq n-1$, has the same
properties as the evolution of~the~concentrated state in the preceding
section (part (D)). We obtain that the~evolution of $\psi_2$ is much
slower than the~QM-evolution -- details are in~the~preceding section. We
obtain that the~dispersion of~$\psi_2$ in~$y_i$, $i = 1, \dots, n-1$, is
of~order
$$
\Delta x_2^2 \leq \Delta x_1^2 + \Delta p_1^2 \frac{T^2}{m^2} \big( 1 +
o(\beta_1 T) \big).
$$
On the other hand, there is no control on~$y_n$, $q_n$. After
transforming this back to variables $\Delta x_i$, $\bar x$ etc. we
obtain that the dispersion of
$$
\sum_{i=1}^n (\Delta x_i^2)^2
$$
is of order $\Delta x_2^2$, while $\bar x^2$ is relaxed.

This implies the following behavior of~the $n$-particle system
\linebreak
in~SubQM$_{CRF}$ (assuming $\beta_1 T \ll 1 \ll \beta_3 T$):

(i) The evolution of~the~correlated state during time $T$ gives
the state in~which the~relative positions are small,

(ii) the mean position behaves quantum-mechanically, because
\linebreak $\beta_3 T \gg 1$, so that the~long-time approximation
applies to $y_n^2 \sim \bar x^2$,

(iii) the resulting picture contradicts QM in this, that the~group
of~particles behaves as a~correlated system, i.e. as a~whole, as
a~certain "superparticle" in~the~QM-law, but the~inner dispersion inside
the~group is much smaller than in QM.


\newpage
\head
9. Proposed experiments
\endhead

Tests which can differ between QM and SubQM are based on~the
existence of~the~concentration, resp. correlation effects in SubQM.

The possible subquantum effects depend on~the~value of~the~parameter
$a$, resp. $\tau_0 = 1 / \beta$ characterizing the subquantum model.
The~values of~parameters $L$, $\delta$, $T_0$, $m$, $V$ characterizing
preparation of~the~particles are related to the value of~$\tau_0$
in~part (B) below.

The goal of~tests is to~find a~result implying the~existence of~some
subquantum effect. The~result would be then the~lower estimate for~the~relaxation
time $\tau_0$. The~upper estimate of~$\tau_0$ is another problem not
discussed in~this paper.

In the description of each test we have to specify:
\roster
\item"(i)" preparation of the state of particles,
\item"(ii)" preparation of the beam of particles,
\item"(iii)" type of searched effect,
\item"(iv)" configuration of the screen and of the other measuring devices,
\item"(v)" description of~results indicating presence of a subquantum effect.
\endroster

Description of tests will be given in three parts in which the first
part (A) will be common for all tests.

(A) Preparation of the state of particles (parameters $L$, $\delta$)
and~pre\-paration of~the~beam of~particles (parameters $T_0$, $m$, $V$) --
i.e. (i) + (ii).

(B) Description of~parts (iii)-(v) for~each particular test.

(C) The discussion of possible physical values of parameters $L$,
$\delta$, $T_0$, $m$, $V$ in the relation to possible values
of~$\tau_0 = 1 / \beta$ (including the~parameter $L_{sc}$
and~other parameters describing geometry of~the~screen).

The idea is to look for concentration and~correlation effects, which are
typical for any subquantum model and~which are excluded by QM. The~first
step is to create, by using iterated slits, the~concentrated
or~correlated state of~particles. In~the~concentrated short-time pulse we
have concentration of~three quantities: space position, momentum
and~time position.

(A) The preparation part of any test consists in passing through
iterated slits. Here we shall describe the standard form of~iterated
slits and the possible variants will be described below. Let us assume
that the particles are moving in the~direction of~the~axis $x_3$. Let
$\delta$ denote the radius of~the~hole and~$L$ denote the distance
between slits/holes. Then the~first hole $H_{-1}$ means the solid screen
with the hole at the center
$$
H_{-1} = \{ x \in \Bbb R^3 | \ x_3 = -L, \ x_1^2 + x_2^2 \geq \delta^2  \}
$$
and the second hole will be
$$
H_0 = \{ x \in \Bbb R^3 | \ x_3 = 0, \ x_1^2 + x_2^2 \geq \delta^2  \}.
$$
The true slits will be one-dimensional objects
$$
\align
S_{-1} &= \{ x \in \Bbb R^3 | \ x_3 = -L, \ |x_1| \geq \delta  \}, \\
S_0 &= \{ x \in \Bbb R^3 | \ x_3 = 0, \ |x_1| \geq \delta  \}.
\endalign
$$

We shall consider in details only the first situation -- the holes,
since the~case with slits is similar. The screen will be at
the~distance~$L_{sc}$,
$$
S_{sc} = \{ x \in \Bbb R^3 | \ x_3 = L_{sc} \}.
$$
One can also consider the case with three or more iterated holes (resp.
slits) with the~other hole
$$
H_{-2} = \{ x \in \Bbb R^3 | \ x_3 = -2 L, \ x_1^2 + x_2^2 \geq \delta^2  \},
\text{etc.}
$$

We shall suppose that particles in the~beam move with the velocity
$$
V > 0
$$
in the direction of the axis $x_3$.

We shall suppose also that the beam has a~form of~a~pulse with
the~duration
$$
T_0 > 0.
$$
We shall consider two types of~pulses:
\newline
-- the short-time pulse satisfying $T_0 \ll \tau_0 = 1/\beta$,
\newline
-- the long-time pulse satisfying $T_0 \gg \tau_0$.

In this way we are able to~prepare particles in~the~concentrated state
assuming that $L$ and~$\delta$ are sufficiently small (with respect
to~$V \tau_0$). Preparation of~the~correlated state requires that
$$
\tau_0 \ll T_0 \ll \tau_1.
$$
Thus the concentrated state may exist in~the~form of~both long-time
and~short-time pulses, while the correlated state requires
the~short-time pulse beam.

The last parameter of a particle will be its mass
$$
m > 0.
$$
The simplest form of~the~screen will be the plain $S_{sc}$ with distance
$L_{sc} = L$. If~$L_{sc} / V~\gg \tau_0$ ($\gg \tau_1$, respectively)
or~$L / V~\gg \tau_0$ ($\gg \tau_1$, resp.) then
probably all subquantum effects disappear, in particular, the
concentration and~correlation effects disappear.

The main step in~the~preparation of~the~beam is to~let it pass through
iterated holes (resp. slits): $H_{-1}$, $H_0$. To obtain
the~concentrated state after passing the~last hole, we need
$$
T := L / V \ll \tau_0, \text{ i.e. } \beta T \ll 1. \tag"(9.1)"
$$
We have
$$
\kappa_0^2 = \left( \frac{2}{\hbar} \Delta x_0 \Delta p_0 \right)^2 \cong
8 \frac{m^2}{T^2 \hbar^2} \cdot \Delta x_0^4 + \frac{4}{9} (\beta T)^4
$$
so that the second term is already small. The concentrated state then
needs (together with $\tau \ll \tau_0$) that
$$
\delta^2 = \Delta x_0^2 \ll \frac{T \hbar}{3 m} \ll \frac{\tau_0
\hbar}{3 m}. \tag"(9.2)"
$$
These two conditions are sufficient for~the~creation of~the~concentrated
state by~passing through two iterated holes (or slits).

The~degree of~the~concentration $\kappa$ depends crucially on~$T = L /
V$. E.g., for $T \gg \tau_0$ (i.e. $L \gg V \tau_0$) $\kappa \gg 1$ and
all subquantum effects disappear.

(B) The distance between the~screen and~the~last hole $H_0$ (resp.
slit $S_0$) will be denoted $L_{sc}$. We shall consider two types
of~measurements:
\roster
\item"(i)" screen,
\item"(ii)" detectors.
\endroster

In the first case -- screen -- we measure the density of~observed
particles. The~measurement is a~measurement of~the~position observable.

Usually the~observed density can be decomposed into a~slowly varying
amplitude part and~a~rapidly oscilating part. We have
$$
\rho (x) \cong \bar \rho (x) \sin^2 \phi (x),
$$
where $\phi$ describes the~"rapidly oscilating" part and~$\bar \rho$ is
the~"slowly varying" component. The~main part density $\bar \rho$ can
be obtained as a~mean value of~$\rho$ over oscilations.

This decomposition into $\bar \rho$ and~$\phi$ is standard
for~interference pictures. If~$\lambda_0$ is the~typical wave-lenght, then
$\bar \rho$ is slowly varying on~distances of~order $\lambda_0$.

In~the~second case, we shall consider certain number of~detectors placed
on~the~screen. Typically we shall consider the~detector as a~hole (or
a~slit) such that particles passing through this hole (or slit) are
registered. We shall consider the~following types of~detectors
$$
\align
H_{sc} \big(x_1^0, x_2^0, r\big) &:= \big\{ x \in \Bbb R^3 |\  x_3 = L_{sc},
\ (x_1 - x_1^0)^2 + (x_2 - x_2^0)^2 < r^2 \big\},\\
S_{sc} \big(x_1^0, r \big) &:= \big\{ x \in \Bbb R^3 |\  x_3 = L_{sc},
\ |x_1 - x_1^0| < r \big\}.
\endalign
$$
The observed number of~particles can be related to~the~particle density
by
$$
Num\, \big[ H_{sc} \big( x_1^0, x_2^0, r \big) \big] = \int_{H_{sc}
(x_1^0, x_2^0, r)} \rho\, dx,
$$
and similarly for $S_{sc} (x_1^0, r)$.

Let the probability that the~particle approaches the~screen
at \linebreak
$H_{sc} (x_1^0, x_2^0, r)$ be $p$, $0 < p < 1$. Let $N_0$ be
the~total number of~particles having passed through the~preparation part
of~the~system. Then the~mean value of~the~number of~particles arriving
at~$H_{sc} (x_1^0, x_2^0, r)$ is
$$
\bar N = N_{all} . p, \ N_{all} = \text{ total number of~particles}
$$
and the~mean quadratic deviation is
$$
\sigma_0 = \sqrt{N_{all}} \sqrt{p (1-p)} = \sqrt{\bar N} \cdot
\sqrt{1-p} \leq \sqrt{\bar N}.
$$
If we assume that $p$ is rather small, we obtain that the~fluctuation
of~the~observed number $N$ of~particles arriving at~$H_{sc} (x_1^0, x_2^0, r)$
is of~order
$$
\sigma_0 \approx \sqrt{N}.
$$
Typical situation is the~following -- there are three detectors
$$
D_0 = H_{sc} (0, 0, r_0), \ D_{\pm} = H_{sc} (\pm x_1^0, 0, r_1),
$$
where $x_1^0 > r_0 + r_1$.

Analogously with slits,
$$
D_0 = S_{sc} (0, r_0), \ D_{\pm} = S_{sc} (\pm x_1^0, r_1), \
x_1^0 > r_0 + r_1.
$$

The evolution after the last hole (slit), i.e. on~$0 < x_3 < L_{sc}$, is
the~following. Let $\Delta x_0^2$, $\Delta p_0^2$ and
$\kappa_0 = \Delta x_0 \Delta p_0 2 / \hbar$ be parameters
at~$x_3 = 0$, i.e. after passing the~last
preparation hole (slit). Then denoting by
$$
\widetilde{\Delta x}_{sc}^2 := \Delta x_0^2 + \Delta p_0^2
\frac{T_{sc}^2}{m^2}, \ T_{sc} = L_{sc} / V,
$$
the "pure" dispersion and~by
$$
\widetilde{\Delta p}_{sc}^2 := \frac{\hbar^2}{4 \widetilde{\Delta
x}_{sc}^2}
$$
the corresponding Heisenberg's dual quantity, we obtain the~"real"
dispersion
$$
\Delta x_{sc}^2 = \Delta x_0^2 + \Delta p_0^2 \frac{T_{sc}^2}{m^2}
+ \frac{4}{9} \widetilde{\Delta p}^2 \cdot \frac{T_{sc}^2}{m^2} \cdot
(\beta T_{sc})^4.
$$
This is rewriting of~formula (7.10)
$$
\Delta x_1^2 = \Delta x_0^2 + \Delta p_0^2 \frac{T^2}{m^2}
+ \frac{1}{9} (\beta T)^6 \frac{\hbar^2}{\Delta p_0^2 (\beta T)^2 +
\Delta x_0^2 m^2 \beta^2}.
$$

This shows that the~conditions for subquantum effects are the~following
$$
T_{sc} := L_{sc} / V \ll \tau_0, \ \text{i.e. } \beta T_{sc} \ll 1,
\tag "(9.3)"
$$
and
$$
\kappa_0 = \frac{2}{\hbar} \Delta x_0 \Delta p_0 \ll 1.
\tag{9.4}
$$
The last condition expresses the fact that the~prepared state must be
concentrated.

Now we can describe proposed experiments. In~the~situation \linebreak
of detectors,
it is reasonable (for obtaining stable results), but not necessary,
to~assume that the~dimensions of~detectors are larger than
the~wave-length of~oscilations
$$
r_0, r_1 \gg \lambda_0. \tag"(9.5)"
$$

(Exp.\,1). In the situation described above assuming that conditions
(9.1)-(9.4) are satisfied, we obtain in~the~"screen-like" situation
the~following relation
$$
\Delta x_{sc}^2 \ll \Delta x_{sc}^{2 (QM)}, \tag"(9.6)"
$$
where $\Delta x_{sc}$ is the~observed dispersion of~the~position
on~the~screen and~$\Delta x_{sc}^{(QM)}$ is the~dispersion expected
by~QM.

This is a~direct manifestation of~the~concentration effect. In~this
case:
\roster
\item"(iii)" Type of the searched effect: the~slower dispersion than
in~QM,
\item"(iv)" configuration: the density on the~screen,
\item"(v)" indication of~the~subquantum effect: inequality (9.6) --
a~small dispersion.
\endroster

(Exp.\,2). Here the situation is the same as in~Exp.\,1, but instead
of~observing the~density on~the~screen we use three detectors $D_0$,
$D_{\pm}$ (in~the~hole or slit variations).

Let us denote by~$N_0$, $N_{\pm}$ the~number of~particles observed
at~detectors $D_0$, $D_{\pm}$ and~let $N_0^{(QM)}$, $N_{\pm}^{(QM)}$ be
corresponding numbers predicted by~QM. Then, assuming (9.1)-(9.5) we have
$$
\frac{N_{\pm}}{N_0} \ll \frac{N_{\pm}^{(QM)}}{N_0^{(QM)}},
$$
or more precisely (but similarly)
$$
\frac{N_{\pm}}{N_{all}} \ll \frac{N_{\pm}^{(QM)}}{N_{all}}.
$$
This expresses the~fact that the dispersion in~the~subquantum situation is
slower than QM-dispersion.

A still more stable indication is expressed by~the~inequality
$$
\frac{N_+ + N_-}{N_0 + N_+ + N_-} \ll \frac{N_+^{(QM)} +
N_-^{(QM)}}{N_0^{(QM)} + N_+^{(QM)} + N_-^{(QM)}}. \tag"(9.7)"
$$
In this case
\roster
\item"(iii)" searched effect: small dispersion,
\item"(iv)" configuration: three detectors $D_0$, $D_{\pm}$,
\item"(v)" indication: inequality (9.7) -- small dispersion.
\endroster

(Exp.\,3). This experiment is proposed for testing possible fluctuation
of~the~basic parameter of~subquantum models -- the~quantity $\tau_0$ --
the relaxation time. This quantity is expressed as a~function
of~the~basic parameter $a$ and the~mass $m$ of~a~particle by
$$
\tau_0^2 = am.
$$
The parameter $a$ describes, in a~sense, the~"density" of~space-like
objects in~SLO-vacuum.

We have assumed that this "density" (and hence also parameter $a$) is
constant with respect to~the~time.

But, by the proper physical idea of~SLO-vacuum, this (and the "density")
is a~dynamical property and it is reasonable to~assume that there may
be fluctuation of~this quantity with respect to~time.

Of course, these fluctuations are significant only on~short time
intervals. Thus it is necessary to~use short-time pulses. Let us
suppose that there are $i = 1, \dots, I$ pulses of~particles, each
of~the~duration $T_0$, where
$$
\beta T_0 \ll 1. \tag"(9.8)"
$$
In~these pulses we have corresponding quantites
$$
N_{all}^i, \ N_0^i, \ N_\pm^i \text{ for } i = 1, \dots, I.
$$
Let the mean values be denoted by
$$
\bar N_0 = I^{-1} \sum N_0^i, \ \bar N_\pm = I^{-1} \sum N_\pm^i.
$$
In the case of~negligible fluctuations, we have the~mean square
deviation satisfying the~inequality mentioned above
$$
\sigma_0 := \left[ I^{-1} \sum_{i=1}^I \big( N_+^i + N_-^i - \bar N_+ - \bar
N_- \big)^2 \right]^{1/2} \leq \left[ \bar N_+ + \bar N_- \right]^{1/2}.
$$
This corresponds to the inequality $\sigma (N) \leq \sqrt{\bar N}$ for
$N = N_+ + N_-$.

In the case of fluctuating value of~$\beta$ we can assume that the~mean
square deviation of~the~quantity $N_+^i +  N_-^i$ will be larger due
to~the change of~$\beta$ and~not only due to~the~standard statistical
deviation. So that we look for satisfaction of~the~opposite
inequality
$$
I^{-1} \sum_{i=1}^I \big( N_+^i + N_-^i - \bar N_+ - \bar N_- \big)^2 > \bar N_+
+ \bar N_-. \tag"(9.9)"
$$
In this experiment it is necessary to~consider the~random quantity
$$
N = N_+ + N_-
$$
with the sample values
$$
N_i = N_+^i + N_-^i, \ i = 1, \dots, I,
$$
because there may exist another subquantum effect -- the~correlation
effect -- which typically gives large fluctuations of~$N_+$
and~$N_-$, but smaller fluctuation of~$N_+ + N_-$ (see below).
Of~course, realization that (9.9) is satisfied needs also
a~reasonably large number $I$ of~pulses. This is a~standard statistical
argument relating $I$ to~the~gap in~inequality (9.9).

We have
\roster
\item"(iii)" searched effect: fluctuation of the~parameter $a$ (resp.
the~density of~SLO's)
\item"(iv)" confiuration: three detectors $D_0$, $D_\pm$,
\item"(v)" indication: inequality (9.9) -- large quadratic deviation
\linebreak
of~$N_+ + N_-$.
\endroster

(Exp.\,4). The basic correlation effect. The~correlation effect uses
the~(hypothetical) correlation between particles in~the~same pulse.

If there are $I$ pulses, $i = 1, \dots, I$, in~each there is a~resulting
density $\rho_i (x)$, $x = (x_1, x_2)$, which is decomposed as
$$
\rho_i(x) \cong \bar \rho_i (x) . \sin^2 \phi_i (x),\ i = 1, \dots, I,
$$
into a~slowly varying part $\bar \rho_i$ and~a~rapidly oscilating part
$\sin^2 \phi_i (x)$.

The total observed density is
$$
\rho (x) \cong \bar \rho (x) . \Phi (x),
$$
where $\bar \rho = \sum \bar \rho_i $ and $0 \leq \Phi \leq 1$ ($\Phi$
is the rapidly oscilating part of~$\rho$).

There are three time intervals
$$
\align
T_0 &= \text{ duration of~the~pulse},\\
T &= L / V,\\
T_{sc} &= L_{sc} / V,
\endalign
$$
and they should satisfy
$$
\align
T_0 &\ll \tau_0,\\
\tau_0 &\ll T \ll \tau_1, \tag"(9.10)"\\
\tau_0 &\ll T_{sc} \ll \tau_1.
\endalign
$$
Thus the length $L_0 = V T_0$ has to satisfy
$$
L_0 \ll V \tau_0 \ll L, \ L_{sc} \ll V \tau_1.
$$

The correlation effect consists in the~following behavior: particles
in~a~pulse behave collectively like one group. This is true in~a~certain
approximation where (9.10) is satisfied.

In consequence of this group-like behavior we can assume (see Sect.\,8)
that
$$
\bar \rho_i (x) \approx \exp \left\{ - \frac{1}{2} \frac{|x -
x_0^i|^2}{r_0^2} \right\}, \ x = (x_1, x_2),
$$
where
$$
x_0^i = (x_{01}^i, x_{02}^i)
$$
is a~center of~this "Gaussian" wave packet and~$r_0$ is its approximate
radius.
The~correlation effect says that centers $x_0^i$ are far from each other
(relatively to~$r_0$), i.e. that
$$
|x_0^i - x_0^j| \gtrsim r_0
$$
at least for many couples $i \not= j$.

In QM any particle in~any pulse is independent on~other particles
and~by~the~standard statistics of~fluctuations we have
for~the~dispersion of~centers
$$
I^{-1} \sum |x_0^i - \bar x_0|^2 \ll r_0^2,
$$
where $\bar x_0 = I^{-1} \sum x_0^i$.

The inequality indicating the~correlation effect is the following
$$
I^{-1} \sum |x_0^i - \bar x_0|^2 \gtrsim r_0^2.
\tag{9.11}
$$
In fact, it is not simple to~identify centers $x_0^i$ (for example
by~re\-gistration of~the~density after each pulse) -- if it is possible
to~decompose reasonably $\bar \rho$ into $\sum \bar \rho_i$, it means
that the~correlation effect takes place. In~the~QM situation we should
have
$$
\bar \rho \approx \exp \left\{ - \frac{1}{2} \frac{|x - \bar x_0|^2}{r^{'2}_0}
\right\},
$$
where $r_0 \lsim r'_0$.

Thus the effect consists in~the~fact that $\bar \rho$ looks like a~sum
of~Gaussians rather that a~certain one Gaussian. Hence experiment
requires a~small number of~pulses:
$$
2 \leq I \leq 8.
$$
We have
\roster
\item"(iii)" searched effect: the group-like behavior of~short pulses,
\item"(iv)" configuration: the~screen, the~short-time pulses,
\item"(v)" indication: inequality (9.11) or a~decomposition of~$\bar \rho$ into
more than one Gaussian.
\endroster

(Exp.\,5). This is also an~experiment looking for correlation effect
using short-time pulses. But instead of~a~screen as in~Exp.\,4 we shall
use detectors $D_0$, $D_{\pm}$.

Let us denote by $N_0^i$, $N_{\pm}^i$ the~number of~particles
of~the~$i$-th pulse arrived at~the~detector $D_0$, resp. $D_{\pm}$. Then
we can calculate the~mean values
$$
\bar N_0 = I^{-1} \sum N_0^i, \ \bar N_{\pm} = I^{-1} \sum N_{\pm}^i.
$$
At first we shall assume that the~total number of~particles in~each
pulse is the~same,
$$
N_{all}^1 = N_{all}^2 = \dots = N_{all}^I.
$$
Let us denote by $\sigma (N_{\pm})$ the mean quadratic deviation
$$
\sigma (N_{\pm}) := \left[ I^{-1} \sum_{i=1}^I \big( N_{\pm}^i - \bar
N_{\pm} \big)^2 \right]^{1/2}
$$
of the (sample) quantities $N_{\pm}^i$. Then by the statistical argument
mentioned above and~using the~basic QM fact stating that all particles
are mutually independent one has the~estimate
$$
\sigma (N_{\pm}) \lsim \sqrt{\bar N_{\pm}}.
$$
This estimate depends on~the~assumption that $I$ (the~number
of~trials-pulses) is sufficiently large. But this is a~standard property
of~all statistical assertions.

Thus the indication of~the~subquantum correlation effect is
fulfillment of~the~opposite inequality (inequalities)
$$
\frac{\sigma(N_{\pm})}{\sqrt{\bar N_{\pm}}} > 1. \tag{9.12}
$$
The gap in~this inequality has to~be considered in~relation with
the~number $I$ of~trials-pulses.

Let us assume that quantum mechanically
$$
\bar N_+ \approx \bar N_-,
$$
which happens in~the~situation when detectors $D_+$ and~$D_-$ are placed
symmetrically. Then, following QM, we have the~standard deviations
of~the~random quantities $N_{\pm}$
$$
N_{\pm}^i \cong \bar N_{\pm} \pm \sqrt{\bar N_{\pm}},\ i = 1, \dots, I,
$$
where $\bar N_{\pm} = I^{-1} \sum N_{\pm}^i$ are the mean values.
Let us define
$$
\bar N := \sqrt{\bar N_+ \bar N_-}.
$$

From
$$
\left| \frac{N_{\pm}^i}{\bar N_{\pm}} - 1 \right| \lsim
\frac{1}{\sqrt{\bar N_{\pm}}}
$$
we obtain
$$
\left| \frac{N_+^i}{\bar N_+} - \frac{N_-^i}{\bar N_-} \right| \lsim
\frac{1}{\sqrt{\bar N_-}} + \frac{1}{\sqrt{\bar N_+}} =
\frac{\sqrt{\bar N_+} + \sqrt{\bar N_-}}{\bar N}
$$
and then
$$
\big| N_+^i \bar N_- - N_-^i \bar N_+ \big| \lsim \bar N \left(
\sqrt{\bar N_+} + \sqrt{\bar N_-} \right).
$$
Using $\sqrt{\bar N_+} + \sqrt{\bar N_-} \cong 2 \sqrt{\bar N}$ we obtain
the~inequality
$$
\left| N_+^i \frac{\bar N_-}{\bar N} - N_-^i \frac{\bar N_+}{\bar N}
\right| \lsim 2 \sqrt{\bar N}.
$$
Here $\frac{\bar N_-}{\bar N}$ and $\frac{\bar N_+}{\bar N}$ are
correction factors related to~the~possibly non-equlibrated situation
$\bar N_+ \not= \bar N_-$.

In~the~correlation effect it often happens that $N_+^i$ and~$N_-^i$ are
\linebreak
substantially different. This gives the~effect that fluctuations
\linebreak
of~$N_+^i - N_-^i$ are larger than in~QM. Thus the~indicating inequality
is
$$
I^{-1} \sum_i \left| N_+^i \frac{\bar N_-}{\bar N} - N_-^i
\frac{\bar N_+}{\bar N} \right| > 2 \sqrt{\bar N}, \ \bar N :=
\sqrt{\bar N_+ \bar N_-}.
\tag{9.13}
$$
In this case we have
\roster
\item"(iii)" searched effect: correlation inside the~group-pulse,
\item"(iv)" configuration: detectors $D_{\pm}$,
\item"(v)" indication: inequality (9.12) or~(9.13) -- large deviation.
\endroster

(Exp.\,6). This is a variant of~Exp.\,5.

There are two detectors in~the~form of~half-plains
$$
\align
D_+ &= \{(x_1, x_2) | \ x_1 > 0 \},\\
D_- &= \{(x_1, x_2) | \ x_1 < 0 \}.
\endalign
$$
Let $N_{\pm}^i$ be numbers of~particles observed at~the~$i$-th  pulse
in~detectors $D_{\pm}$. Let mean values be
$$
\bar N_{\pm} := I^{-1} \sum_{i=1}^I N_{\pm}^i.
$$
It is possible to define the~corrected numbers
$$
\widetilde N_{\pm}^i := N_{\pm}^i \cdot \frac{\bar N_+ + \bar N_-}{N_+^i +
N_-^i}.
$$
Then we can consider inequalities (9.12) or~(9.13) written for
$\widetilde N_{\pm}^i$ as indicating presence of~subquantum effects.

We have
\roster
\item"(iii)" searched effect: correlation inside a~pulse,
\item"(iv)" configuration: two detectors -- half-plains,
\item"(v)" indication: inequalities (9.12) or~(9.13) for corrected
numbers $\widetilde N_{\pm}^i$.
\endroster
This type of~an~experiment was proposed in~[6] under the~name
"subquantum coherence effect". It is possible to~consider experiments
that are variants of~those already proposed. For~example

(Exp.\,2'). The~indicating inequality may be also
$$
\frac{N_0}{N_{all}} \gg \frac{N_0^{(QM)}}{N_{all}}.
$$

(C) In part C we have to~consider concrete forms of~proposed
experiments. There are two possibilities:
\roster
\item"(C1)" massive particles like electrons or~protons or~neutrons,
\item"(C2)" mass-less particles -- photons.
\endroster
Our theory is purely non-relativistic, so that formulas used in~this
section cannot be directly applied to~photons. But we shall consider
experiments Exp.\,1-Exp.\,6 also for photons by~analogy and~by using
$V = c$, i.e. $T = L / c$, $T_{sc} = L_{sc} / c$ etc.

Of course, all possible results indicating presence of~a~certain
subquantum effect depend on~the~parameters of~considered subquantum
model. Namely on~parameter $a$, or, equivalently, $\tau_0$ or~$\tau_i$
(resp. $\beta$ or~$\beta_i$).
There is no indication how large or small this parameter can be.

(C1). Massive particles. In~this part we shall propose possible physical
values of~parameters. There are different cases corresponding
to~a~possible value of~$\tau_0$.

We need to~satisfy the~following inequalities
$$
T \ll \tau_0, \ \delta^2 \ll \frac{T \hbar}{3 m}.
$$
We shall calculate values for~$m = m_e$, the~mass of~the~electron.
\roster
\item"(i)" If we assume that
$$
\tau_0 \sim 10^{-9} s
$$
and using the~approximate values $\hbar \sim 10^{-34} Js$, $m_e \sim
10^{-30} kg$ we obtain the~possible values
$$
\align
T, T_{sc} &\sim 10^{-10}s,\\
\delta &\sim 10^{-7}m
\endalign
$$
and then
$$
L, L_{sc} \lsim V . 10^{-10} m,
$$
where $V$ is the velocity of~particles. If~$V = 0.3 c$ then
$$
L, L_{sc} \lsim V . 10^{-2} m.
$$
\item"(ii)"
If we assume
$$
\tau_0 \sim 10^{-11} s,
$$
then
$$
\align
T, T_{sc} &\sim 10^{-12} s,\\
\delta &\sim 10^{-8} m
\endalign
$$
and then for $V = 0.3 c$
$$
L, L_{sc} \lsim 10^{-4} m.
$$
The appropriate value of~$T_0$ will be in~both cases
$$
T_0 \lsim 0.1 T.
$$
The appropriate value of~$I$ will be $I \gtrsim 100$ but also $I \gtrsim
10$ may be already significant.
\endroster

(C2). Photons.
All experiments can be considered as optical experiments with photons.
All formulas using the~mass $m$ are more or~less meaningless.

We can consider all proposed experiments with photons, but
\roster
\item"(i)" without formulas containing the~mass $m$, i.e. the~velocity
$$
V=c,
$$
the times $T = L / c$, $T_{sc} = L_{sc} / c$.
\item"(ii)" We cannot calculate $\beta$ (nor~$\tau_0$) from the~parameter $a$,
but we can suppose that there is certain time $\tau_0$, possibly $\tau_0
= \tau_0 (\nu)$, where $\nu$ is frequency of~the~light, which defines
the~relaxation time.
\item"(iii)" We can look for subquantum effects trying different combinations
of~parameters $L$, $L_{sc}$, $\delta$. There are two inequalities which
have to~be~satisfied
$$
T \ll \tau_0, \ T_{sc} \ll \tau_0, \text{ i.e. } L, \ L_{sc} \leq \tau_0 c.
\tag{9.14}
$$
\endroster
In experiments Exp.\,4-Exp.\,6 with pulses we have to~assume that
the~"length" of~the~pulse is sufficiently short,
$$
c T_0 \ll L, \ L_{sc},
\tag{9.15}
$$
where $T_0$ is the~duration of~the~pulse.

The correlation effect is more clear, if the~"length" of~the~pulse is
sufficiently short, so that the~pulse should be as short as possible.

The~values of~parameters:
\roster
\item"(i)" if, say, $\tau_0 = 10^{-9} s$, then $\tau_0 c = 0.3\,m$
and~this seems to~be too long;
\item"(ii)"if, say, $\tau_0 = 10^{-11} s$, then $\tau_0 c = 3\,mm$
and~we can assume $L$, $L_{sc} \lsim 1\,mm$ and for the~pulse we can
suppose that $T_0 = 10^{-12} s$, i.e. $c T_0 \sim 0.3\,mm$;
\item"(iii)" the best way is to look for~the~pulse as short as possible,
say, $T_0 \lsim 10^{-13}s$, i.e. $T_0 c \lsim 0.03\,mm$ and then consider
the~length
$$
L, \ L_{sc} \gg T_0 c.
$$
\endroster
The value of~$\delta$ should be reasonable with respect to~$L$
and~$L_{sc}$.

The advantages of~optical experiments are
\roster
\item"(i)" the possibility of~extremally short pulses,
\item"(ii)" large range of~wave phenomena,
\item"(iii)" good standard detectors.
\endroster

The disadvantage
\roster
\item"(i)" there is not an~explicit subquantum model.
\endroster

A typical optical frequency is
$$
\nu \sim 5. 10^{14} Hz.
$$
Corresponding energy is
$$
E_{opt} = \hbar \, \omega \sim 3. 10^{-19} J
$$
and the~"effective" mass is
$$
m_{opt} = E_{opt} . c^{-2} \sim 3 . 10^{-36} kg.
$$
For $T \sim 10^{-10} s$ we obtain the "effective"
$$
\delta \sim 3 . 10^{-4}.
$$

\newpage

\head
Conclusions
\endhead

We have presented subquantum models that can be useful in~at~least two
directions:
\roster
\item"(i)" new phenomena in~subquantum models,
\item"(ii)" new quantization procedure.
\endroster
The new subquantum effects considered in~this paper are:
\roster
\item"(a)" the concentration effect which says that under certain
conditions particles move almost deteministically during short-time
intervals,
\item"(b)" short pulses of~particles move as a~group.
\endroster
The general subquantum models are based on~the~hypotheses of
\roster
\item"(a)" deterministic quantum model,
\item"(b)" the subquantum medium composed of~space-like objects.
\endroster
The influence of~this subquantum medium on~deterministic quantum
particles is modelled by~(quantum) random forces. There are two models:
independent random forces (the~model SubQM$_{RF}$) and correlated random
forces (the~model SubQM$_{CRF}$).

The short-time behavior of~the~subquantum models contain
concentration and~correlation effects, which are basis of~proposed
experiments distinguishing between QM and~subquantum models.

We have proposed a~new quantization procedure which divides quantization
into two steps.
\roster
\item"(i)"
The first step is to~postulate the~corresponding deterministic quantum
model DetQM. We shall call this step the~{\it \linebreak 0-th quantization}.
Construction of~the~deterministic quantum model is defined uniquely
by~the~classical system, this model is completely local.
The~corresponding theory is the~quantum model containing $\hbar$. This
is something like a~"bare" quantum theory. There are all quantum
probability rules (Feynman's rules), but the~evolution is deterministic,
without spreading out of~wave functions. Heisenberg's uncertainty
principle is completely violated in~this deterministic quantum model.
There are states with an~exact localization in~the~position
and~momentum variables. These states do not spread out and~remain
localized in~the~deterministic quantum model.
\item"(ii)"
The~second step consists in~an~introduction of~the~appropriate
subquantum medium and, moreover, in~representation of~this medium as
a~quantum random force acting on~the~deterministic quantum system.
The~second step is a~dynamical mechanism which assures (approximate)
satisfaction of~Heisenberg's uncertainty principle.
\endroster
There are at~least two areas of~application of~this quantization schema:
\roster
\item"(a)" Quantum gravity,
\item"(b)" renormalization in~QFT.
\endroster

0-th quantization of~a~gravity. With respect to~(i), an~important step
will be construction of~deteministic quantum gravity (even construction
of~deterministic Newtonian quantum gravity would be a~big step
forward). As noted by many theoreticians (e.g. R. Penrose), there is
the~main conflict between locality of~the~General Relativity
and~non-locality of~QT. We think that the~construction of~deterministic
quantum gravity would be possible, since both theories, General
Relativity and~Deterministic QT are local. Thus there will be no
conflict on~locality. On~the~other hand, introduction of~the~subquantum
medium may reflect more closely conditions at~the~Big Bang (or
in~the~early period of~the~Universe).

Renormalization of~QFT. The behavior of~subquantum models on~small
distances is milder than in~the~standard QFT. There is a~hope that
the~QFT will not require (infinite) renormalization and~that
the~subquantum models of~QFT will be finite. (The~finite renormalization
will be, of~course, as useful as before.)

\newpage

\enddocument
\bye